\documentclass[notitlepage,longbibliography,
 amsmath,amssymb,
prfluids,
]{revtex4-2}

\usepackage{dcolumn}% Align table columns on decimal point
\usepackage{bm}% bold math
\usepackage[usenames,dvipsnames]{xcolor}

\usepackage{amsmath}
\usepackage{siunitx}
\usepackage{graphicx}

\usepackage{ulem}
\newcommand{\ab}[1]{\textcolor{Black}{#1}}

\newcommand\abceqn[2]{\refstepcounter{equation}
     \[
     \label{#1}
     #2
     \eqno{\text{(\theequation)}\text{a,b,c}}
     \]
}
\newcommand\abcdeqn[2]{\refstepcounter{equation}
     \[
     \label{#1}
     #2
     \eqno{\text{(\theequation)}\text{a,b,c,d}}
     \]
}

\newcommand{\bendotaxis}{bendotaxis} % have taken the italics away
\newcommand{\xleft}{x_-}
\newcommand{\thetaleft}{\theta_-}
\newcommand{\xright}{x_+}
\newcommand{\thetaright}{\theta_+}
\newcommand{\dd}[2]{\frac{\mathrm{d}#1}{\mathrm{d}#2}}
\newcommand{\ddp}[2]{\frac{\partial #1}{\partial #2}}
\newcommand{\hyspar}{\lambda}
\newcommand{\hysparmax}{\hyspar_{\text{max}} }
\newcommand{\hyspare}{\lambda_e}
\newcommand{\hyspari}{\lambda_{\infty}}
\newcommand{\thetai}{\theta_\pm} %collective term for contact angles
\newcommand{\xmi}{x_\pm} %collective term for menisci

\newcommand{\xlefteq}{X_{-}}
\newcommand{\xrighteq}{X_{+}}
\newcommand{\xpmeq}{X_{\pm}} %collective term for meniscus positions in equilibrium

\begin{document}

\title{Droplet trapping in bendotaxis caused by contact angle hysteresis}

\author{Alexander T. Bradley}
 \altaffiliation[Now at ]{British Antarctic Survey, Cambridge, UK}%Lines break automatically or can be forced with \\
\author{Ian J. Hewitt}%
\author{Dominic Vella}%
\email{dominic.vella@maths.ox.ac.uk}
\affiliation{%
Mathematical Institute, University of Oxford, Woodstock Rd, Oxford, OX2 6GG, United Kingdom
}%


\begin{abstract}
Passive droplet transport mechanisms, in which continuous external energy input is not required for motion, have received significant attention in recent years. Experimental studies of such mechanisms often ignore, or use careful treatments to minimize, contact angle hysteresis, which can impede droplet motion, or even arrest it completely. Here, we consider the effect of contact angle hysteresis on \bendotaxis, a mechanism in  which droplets spontaneously deform an elastic channel via capillary pressure and thereby  move. Here, we seek to understand when contact angle hysteresis prevents \bendotaxis. We supplement a previous mathematical model of the dynamics of \bendotaxis~with a simple model of contact angle hysteresis, and show that this model predicts droplet trapping when hysteresis is sufficiently strong. By identifying the equilibrium configurations adopted by these trapped droplets and assessing their linear stability, we uncover a sensitive dependence of \bendotaxis~on contact angle hysteresis and develop criteria to describe when droplets will be trapped.
\end{abstract}

\maketitle

\graphicspath{{./Figures/}}
\section{Introduction}
%we're interested in controlled movement of small droplets
The transport of liquid droplets on small scales, where surface forces dominate over body forces, occurs in myriad applications, ranging from droplet-based microfluidics~\citep{Squires2005RevModPhys} and medical diagnostics~\citep{Yager2006Nature} to fog harvesting~\citep{Andrews2011Langmuir} and microfabrication~\citep{Srinivasarao2001Science}. In many scenarios, such droplet transport is achieved by active control of the droplet, usually through an applied pressure gradient. However,  there has also been particular interest in passive droplet transport mechanisms, which  do not require a continuous external energy input. Within this category, mechanisms can be further classified into those that exploit a fixed geometry, such as placing droplets in wedges~\citep{Renvoise2009EPL, Reyssat2014JFM} or on cones~\citep{Lv2014PRL, McCarthy2019SoftMatter, Lorenceau1999JFM}, and those that generate motion via deformation of their solid confines. Examples of mechanisms relying on solid deformations include durotaxis~\citep{Style2013PNAS, Bueno2018SoftMatter} --- droplet motion in response to gradients in stiffness of the underlying substrate --- and tensotaxis~\citep{Bueno2017EML} --- droplet motion in response to gradients in strain of the underlying substrate.

\begin{figure}[t]
    \centering
   \includegraphics[width = .95\textwidth]{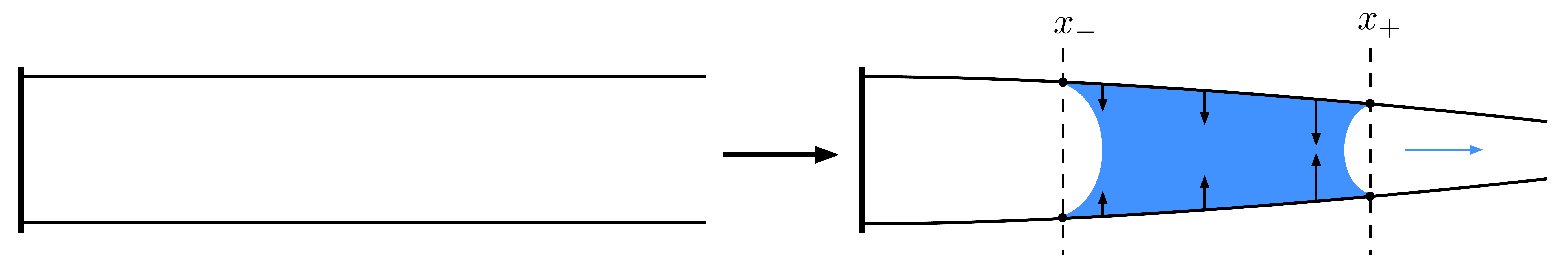}
    \caption{Schematic diagram illustrating the mechanism behind \bendotaxis: an empty, two-dimensional channel with one end clamped and the other free (left panel) experiences a deformation when a liquid droplet that wets the channel walls is introduced (right panel). The resulting deformation (magnitude indicated  by black arrows) is larger at the meniscus closer to the free end ($\xright$) than at the meniscus closer to the clamped end ($\xleft$), creating a pressure gradient that drives the droplet toward the free end (blue arrow). }
    \label{fig:Mechanism}
\end{figure}

% Description of bendotaxis
One example of an entirely passive droplet-driven motion is offered by interactions of droplets with bendable fibres or plates, which has been termed `bendotaxis' \citep{Bradley2019PRL}. The essential mechanism of \bendotaxis~is that surface tension forces associated with droplets cause the elements to bend thereby creating a tapering that propels the droplets. This tapering relies on anisotropic channel clamping conditions e.g.~clamped at one end and free at the other. While examples of this bending-induced droplet self-propulsion have been studied for droplets trapped between cylindrical hairs \cite{Duprat2012Nature,Wang2015PNAS}, it is easier to understand the interaction between bending and capillary pressure for a droplet trapped within a deformable channel; Figure~\ref{fig:Mechanism} elucidates the mechanism behind \bendotaxis~in this case. The negative pressure associated with a wetting droplet introduced into the channel results in an inwards deflection of its walls. Owing to the anisotropy in clamping conditions, the resulting deformation is larger at the meniscus closer to the free end (referred to as $\xright$) than at the clamped end ($\xleft)$. The pressure is therefore more negative at $\xright$ than at $\xleft$; the resulting pressure gradient drives the droplet towards the free end. In the absence of contact angle hysteresis, and, provided that the walls do not touch, this motion will continue until the droplet reaches the free end. (Note that this mechanism, albeit with a positive Laplace pressure and outwards deformation,  also results in non-wetting droplets spontaneously moving in the same direction; here we consider only wetting droplets for simplicity.)

% In many of these passive mechanisms, the motion can be scuppered by contact angle hysteresis
The growing list of passive droplet transport mechanisms described above is the result of intensive investigation, particularly experimentally.  Naturally, the main focus of these studies is to gain an understanding of the physics that gives rise to the force imbalance and thus droplet motion. However, since this force imbalance depends sensitively on the meniscus curvature, and hence the contact angle of the droplet, it is also sensitive to contact angle hysteresis --- the asymmetry between advancing and receding contact angles that results from local liquid pinning on inhomogeneities in the surface~\citep{deGennes2004}. In practice, hysteresis is often carefully controlled (e.g.~by using `slippery'  surfaces that are close to hysteresis-free~\citep{McCarthy2019SoftMatter}). Alternatively, at the theoretical level it is usually treated in a static fashion~\citep{Lv2014PRL}, or neglected entirely. In the scenarios where these mechanisms are intended to be exploited, however, conditions cannot always be carefully controlled, and some hysteresis will be present; it is therefore of practical importance to understand the influence of hysteresis on these droplet transport mechanisms. The worst case scenario from the  perspective of  droplet transport is that contact angle hysteresis completely arrests the motion, as has been shown to be possible recently for droplets in tapered channels~\citep{Prakash2008Science, Bush2010AdvCollIntsci} (though these studies also showed that contact angle hysteresis can be used to develop a tweezer for droplets, demonstrating that contact angle hysteresis can also be usefully exploited if properly understood).

%here we consider the effect of contact angle hysteresis on one method of droplet motion, with a particular focus on understanding when contact angle hysteresis is sufficient to trap the droplet indefinitely. We consider \bendotaxis, which relies on gradients in bending stiffness.describe mechanism. Say that, although it works for both wetting and non-wetting droplets, here we'll focus on wetting droplets (where both contact angles on the same side of 90deg. We supplement this with simple (simplest?) model of contact angle hysteresis.  Need to mention droplets continue until they reach the end of the channel always under this model. Here we supplement this with a simple model of contact angle hysteresis

In this paper, we focus on the effect of contact angle hysteresis on \bendotaxis. In \bendotaxis, the droplet motion results from the (self-induced) tapering of the channel; since droplets in tapered channels with externally imposed channel tapering can be trapped part-way along the channel as a result of contact angle hysteresis, we might expect a similar scenario in the \bendotaxis~mechanism. This leads to the two main questions that we aim to answer in this paper: can contact angle hysteresis prevent droplets from self propelling along deformable channels by \bendotaxis? And, if so, when does this hysteresis-induced `trapping' occur? 

%paper structured as follows
This paper is structured as follows. In \S\ref{S:Model} we provide a brief outline of the mathematical model of \bendotaxis~described by~\cite{Bradley2019PRL}, which we supplement with a simple model of dynamic contact angle hysteresis. In non-dimensionalizing this model, we identify four key dimensionless parameters: a channel bendability, a contact angle hysteresis parameter, a dimensionless droplet volume and a dimensionless initial droplet position. The remainder of the paper is dedicated to understanding if and, if so, when (i.e.~in which regions of this four dimensional parameter space) is contact angle hysteresis sufficiently strong to prevent droplets from self-propelling along the channel. In \S\ref{S:Numerics}, we present numerical solutions of the governing equations; these solutions confirm that, when hysteresis is sufficiently strong, droplets may get trapped and offer insight into how the system parameters affect whether droplets will be trapped. Following this,  \S\ref{S:Equilibria} concerns the trapped configurations of droplets --- i.e.~the equilibria of the system. We map out the locations of these equilibria in parameter space and assess their linear stability. In \S\ref{S:MobileDroplets}, we address the central question of the paper:  when does contact angle hysteresis prevent droplets from self-propelling along deformable channels? By making an approximation that droplets that are ultimately trapped do not move appreciably from their initial positions, we re-purpose the equilibrium maps developed in \S\ref{S:Equilibria} to describe whether droplets of given parameters will be trapped or not. Finally, in \S\ref{S:Conclusions}, we summarize our findings and discuss possible directions for further investigation.

\section{Mathematical Model}\label{S:Model}
%In this section, we describe the mathematical model, based on the \bendotaxis~model described by~\cite{Bradley2019PRL} with dynamically evolving contact angles. We then couple this model to a simple model of contact angle hysteresis via the dynamic contact angles. 

\begin{figure}[t]
\centering
\includegraphics[width = 0.7\textwidth]{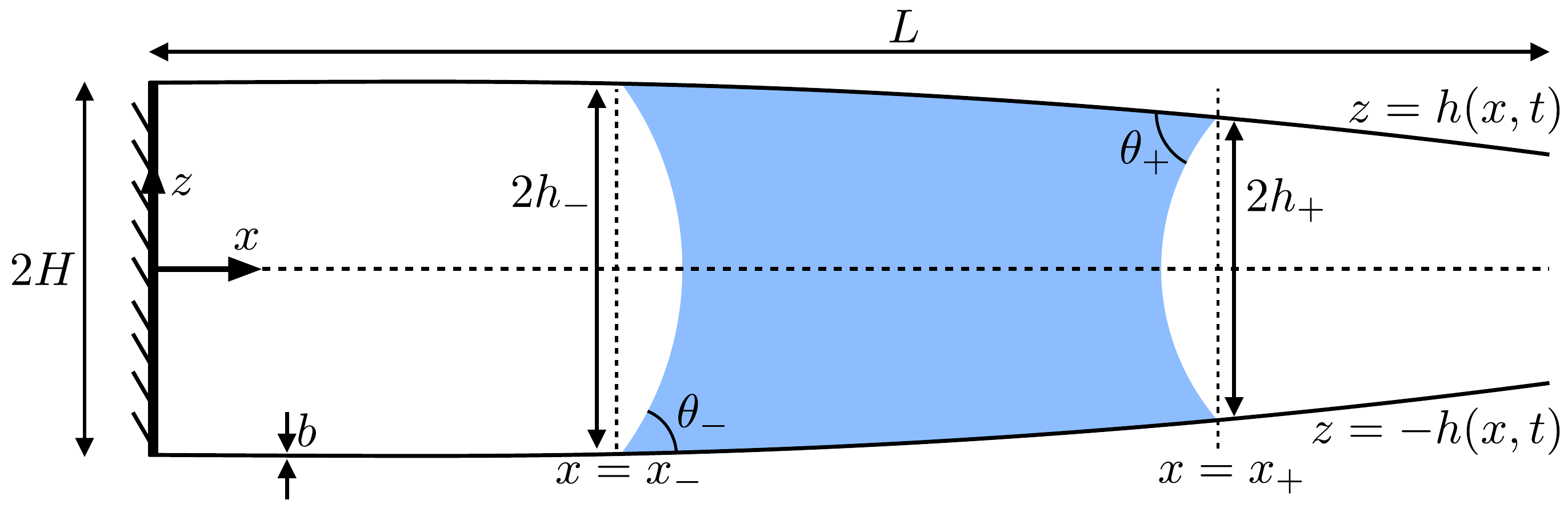}
\caption{(a) Schematic diagram of a droplet in a two-dimensional channel consisting of two flexible walls of thickness $b$ and a rigid end. The channel has undeformed wall separation $2H$. The menisci contact the walls at perpendicular distances $x = \xleft(t)$ and $x = \xright(t) $ from the clamped end of the channel, with contact angles $\theta = \theta_-(t)$ and $\theta = \theta_+(t)$, respectively.}\label{fig:Schematic}
\end{figure}

We consider the setup shown in Figure~\ref{fig:Schematic}: a channel bounded by two narrow, flexible beams of thickness $b$, length $L$, density $\rho_s$\ab{,} and Young’s modulus $E$, are clamped parallel to one another at a distance $2H$ apart, at one end of the beams. This clamped end defines the $z$-axis, and the axis of the channel (parallel to the undeformed beams) defines the $x$-axis; $z=0$ is defined to be the centre of the undeformed channel, while the deformed channel walls lie at $z=\pm h(x,t)$. \ab{Here we consider only behaviour in the $(x,z)$-plane, but assume for simplicity that the channel is relatively narrow (width much smaller than the channel length $L$) in the direction into the page.}

The channel contains a droplet of liquid of viscosity $\mu$ and density $\rho$. The droplet has (two-dimensional) volume $\Omega$, and makes a liquid bridge between the channel walls, wetting them over the region $\xleft(t) < x < \xright(t)$ (we assume that the droplet-channel system is symmetric about the centre-line $z = 0$, so that this contact point is identical on both sides of the droplet). The droplet makes a contact angle $\theta_{\pm}(t)$ at the menisci located at $x_{\pm}$, respectively; it is through the dynamically evolving contact angles $\theta_{\pm}$ that we include contact angle hysteresis in our model. 

\subsection{Fluid flow model}
We assume that the drop is long and thin, $\Omega/H^2\gg1$, so that lubrication theory~\cite{Leal2007} applies. Within this framework, the local conservation of mass combined with the kinematic boundary condition at the channel walls ensures that the droplet pressure $p(x,t)$ and channel half-width $h(x,t)$ satisfy Reynolds' equation \cite{Leal2007}\ab{:}
\begin{equation}\label{E:Model:Reynolds}
\ddp{h}{t} = \frac{1}{3\mu}\ddp{}{x}\left(h^3 \ddp{p}{x}\right).
\end{equation} The pressure within the liquid, $p(x,t)$ is coupled to the channel shape, $h(x,t)$, as we shall discuss shortly. However, we first discuss the boundary conditions on pressure that are appropriate.

The pressure at the droplet menisci depends on the meniscus shape. For small Bond number droplets, $\rho g H^2 / \gamma \ll 1$, the effect of hydrostatic pressure on the droplet can be neglected; in particular, the menisci are minimal surfaces, i.e.~they are approximately arcs of circles with curvatures
\begin{equation}\label{E:Model:curvatures}
\kappa_{\pm} = -\frac{\cos \thetai}{h(x = \xmi,t)}.
\end{equation}
The pressure boundary conditions imposed on~\eqref{E:Model:Reynolds} are therefore
\begin{equation}\label{E:Model:LaplacePressure}
p =- \frac{\gamma \cos \thetai}{h} \qquad \text{at}~x = \xmi,
\end{equation}
where $\gamma$ is the surface tension of the air--liquid interface.

Droplet motion is driven by the pressure difference along the droplet. In typical laboratory conditions, the time-scale of evaporation is significantly longer than the time-scale of droplet motion~\cite{Bradley2019PRL}.  Evaporation can therefore be ignored and the flux of fluid through the menisci must balance that caused by motion, giving the kinematic conditions
\begin{equation}\label{E:Model:Kinematic}
\dd{\xmi}{t} = -\left.\frac{h^2}{3\mu}\ddp{p}{x}\right|_{x = \xmi}.
\end{equation}

\subsection{Beam deflection model}

To couple the pressure within the droplet to the shape of the channel walls, we use linear beam theory~\citep{Howell2009}. This theory is valid provided that the beams are thin ($b \ll L$)  and undergo small deformations in comparison with their length (which is guaranteed if $H \ll L$, and is consistent with our use of lubrication theory in the fluid). In this framework, the shape of the channel wall, $h(x,t)$, satisfies the Euler--Bernoulli equation
\begin{equation}\label{E:Model:EulerBernoulli}
B\ddp{^4 h}{x^4} = q(x,t),
\end{equation}
where $B = Eb^3 /12$ is the bending stiffness of the channel walls (independent of Poisson's ratio because the walls are narrow~\citep{Audoly2010elasticity}) and $q(x,t)$ is the applied load, which is equal to the droplet pressure in the wetted portion of the beam and zero otherwise, i.e.
\begin{equation}\label{E:Model:DropletPressure}
q(x,t) =\left\{ \begin{array}{l l}
0 & \qquad \text{for}~ 0 < x < \xleft(t),\\
p(x,t) & \qquad\text{for}~ \xleft(t)< x < \xright(t),\\
0 & \qquad \text{for}~ \xright(t) < x < L.
\end{array}\right.
\end{equation}

(Note that in using the static beam equation \eqref{E:Model:EulerBernoulli}, we have neglected the wall inertia and weight \ab{since ref}~\cite{Bradley2019PRL} showed that, in typical experimental conditions, wall inertia and the weight of both the channel wall and droplet were both negligible.

By combining~\eqref{E:Model:Reynolds},~\eqref{E:Model:EulerBernoulli},  and~\eqref{E:Model:DropletPressure}, we can eliminate the droplet pressure to give a system of partial differential equations (PDEs) for the channel half-width\ab{:}
\begin{align}
0 &= \ddp{^4 h}{x^4} & &0 < x <\xleft(t)\ab{,} \label{E:Model:CombinedEq1} \\
\ddp{h}{t} &=  \frac{\color{Green}{B}}{3\mu}\ddp{}{x}\left(h^3 \ddp{^5 h}{x^5}\right)   & &\xleft(t) < x <  \xright(t),\label{E:Model:CombinedEq2}\\
\qquad \qquad 0 &= \ddp{^4 h}{x^4} &  &  \xright(t) < x < 1.\label{E:Model:CombinedEq3}
\end{align}

To proceed further, we require boundary conditions. We note first that combining \eqref{E:Model:DropletPressure} with~\eqref{E:Model:LaplacePressure} and~\eqref{E:Model:EulerBernoulli} gives 
\begin{equation}    \label{E:Model:PressureExplicit}
%\left.\frac{\partial^4h}{\partial x^4}\right|_{x=x_\pm^\mp}=-\frac{\gamma\cos\theta_\pm}{B}h(x_\pm,t)^{-1}
\left[\ddp{^4h}{x^4}\right]_{x = x_{\pm}} = \mp \frac{\gamma\cos\theta_\pm}{B}h(x_\pm,t)^{-1}
\end{equation}
where square brackets denote the jump in a quantity across the meniscus denoted in the subscript, e.g.~for $\xright$:
\begin{equation*}
    \left[f\right]_{\xright} = \lim_{x \downarrow \xright} f(x) -  \lim_{x \uparrow \xright} f(x).
\end{equation*} In contrast to the discontinuity in the fourth derivative of $h(x,t)$ at the menisci, we assume that $h$ and its first three derivatives (corresponding to the beam slope, moment, and shear force, respectively) are continuous across the menisci, i.e. 
\abcdeqn{E:Model:ContinuityBC}{
\left[h\right]_{\xmi} =0, \quad \left[\ddp{h}{x}\right]_{x_{\pm}} =0, \quad \left[\ddp{^2 h}{x^2}\right]_{x_{\pm}} = 0, \quad \left[\ddp{^3 h}{x^3}\right]_{x_{\pm}}=0.}
In~\eqref{E:Model:ContinuityBC} we have ignored the line force from surface tension. The validity of our neglect of the line force may be determined by considering the net force exerted by the droplet on the beam: with the line force included, the droplet pressure $p \sim \gamma \sin \theta_+ \delta(x - x_+) + \gamma \cos \theta_+ / H $, where $\delta$ is a Dirac $\delta$-function, and the total force on the beams is
\begin{equation*}\label{E:Model:TotalPressure}
\int_{x_-}^{x_+}p~\mathrm{d}x \sim \gamma \sin \theta_+ + \gamma \cos \theta_+ \frac{L}{H}.
\end{equation*}
Comparing the contribution to the total force from the line force (first term above) with the large scale contribution from surface tension (second term) demonstrates that the former can be neglected provided that $\tan \theta_{\pm} \ll L/H$. This holds for the very slender channels considered experimentally by reference~\cite{Bradley2019PRL}, provided that the contact angle is not close to $90^\circ$.

Having considered boundary conditions at the edge of the droplet, we must also impose boundary conditions at the two dry ends of the beams. We impose clamped boundary conditions at $x = 0$:
\begin{equation}\label{E:Model:ClampedBC}
h = H\quad \text{and} \quad  \ddp{h}{x} = 0 \quad \text{at}~x = 0,
\end{equation}
and assume that at their far end ($x=L$) the beams are free --- they are not subject to any moment or shear --- so that
\begin{equation}\label{E:Model:FreeBC}
\ddp{^2 h}{x^2} = 0 \quad \text{and} \quad  \ddp{^3 h}{x^3}= 0   \quad \text{at}~x = L.
\end{equation} 
In making this `free end' assumption,  we are neglecting the possibility that the ends of the beams may touch, for example if the droplet surface tension is sufficiently strong. As we shall see, those droplets that become trapped typically do so close to their initial position, and while the channel wall displacements remain small, making this a reasonable assumption. The case in which the ends touch has been considered by ref~\citep{Bradleyphdthesis}.

%Note also that when surface tension is sufficiently strong, we expect that the beams may touch during droplet motion. In this case, the boundary conditions~\eqref{E:Model:FreeBC} will no longer hold, but we do not consider this scenario here %i.e. for the trapping of interest, we only need free BC

The asymmetry in boundary conditions between \eqref{E:Model:ClampedBC} and \eqref{E:Model:FreeBC} --- clamped at one and, and free at the other --- mean\ab{s} that, for a given imposed force, a larger deflection is observed towards the free end of the channel (i.e.~the channel is effectively `softer' towards the free end, even though the bending stiffness is constant).  This asymmetry is a crucial part of the mechanism that drives \bendotaxis.

\subsection{Contact Angle Hysteresis}\label{S:Model:ModellingHysteresis} 

\begin{figure}[t]
    \centering
    \includegraphics[width = \textwidth]{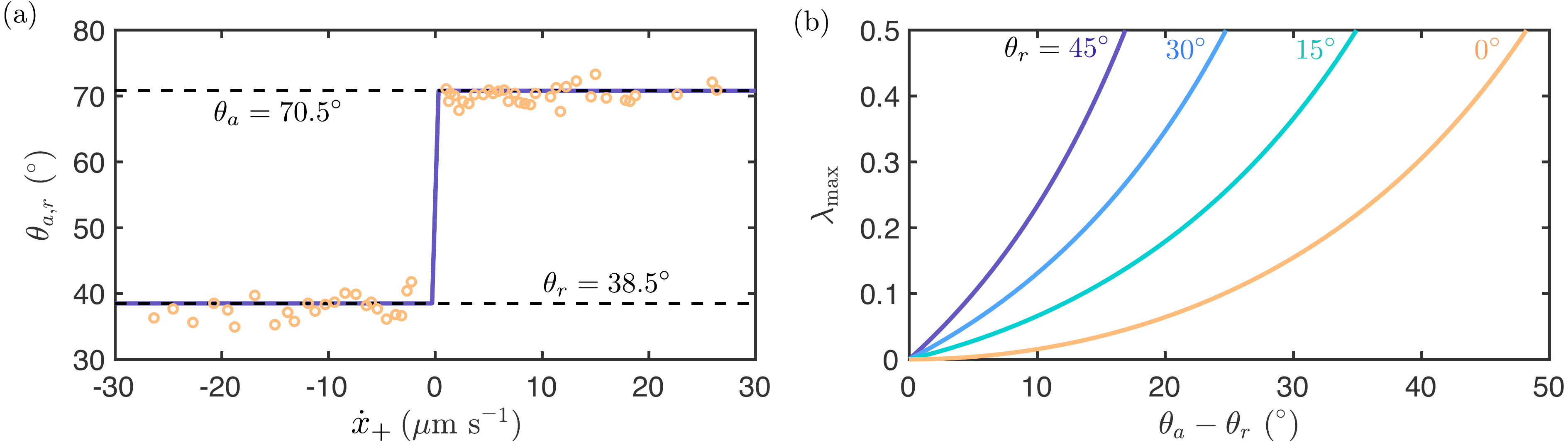}
    \caption{(a) Plot of the dynamic contact angle law~\eqref{E:Model:theta_+_conditions} relating $\theta$ to the contact line speed $\dot{x}_+$, fitted to \ab{example} experimental data from \citep{Petrov1991} \ab{for a water droplet on a polyethylene terephthalate substrate with $\theta_a = 70.5\si{\degree}$, $\theta_r = 38.5\si{\degree}$} (points reproduced from their Figure 6).
 (b) Plots of the hysteresis parameter $\hysparmax$, defined in~\eqref{E:Model:hysteresis_measures}, as a function of $\theta_a - \theta_r$ for various advancing angles $\theta_r$ (as indicated by the labels).}
    \label{fig:hysteresis_graph}
\end{figure}

A key parameter in our model of dynamic \bendotaxis~is the contact angle that each meniscus makes with the beam, denoted $\theta_\pm$. While the energetically preferred equilibrium value of the contact angle is determined by a balance between the surface energies of the three phases that meet at the contact line, it is also known that this value can be modified by the presence of microscopic defects to give contact angle hysteresis~\cite{Joanny1984,deGennes2004}. Moreover, even in the absence of defects, hydrodynamic effects mean that the contact angle observed in dynamic scenarios may differ substantially from its equilibrium value~\cite{Snoeijer2013}.

Many different models for contact angle hysteresis and for the dynamic contact angle have been proposed~\cite{Snoeijer2013}. We adopt perhaps the simplest possible model that allows different advancing and receding contact angles with the key feature that a jump in the contact angle occurs at zero meniscus velocity. In particular, we assume that: (i) the droplet--channel system has intrinsic static advancing and receding contact angles, $\theta_a\geq\theta_r$, respectively; (ii)  a stationary interface may take any contact angle $\theta_r \leq \theta \leq \theta_a$; (iii) a dynamic meniscus has a constant contact angle equal to $\theta_a$ if the meniscus is advancing (liquid-invading-vapour) or $\theta_r$ if the meniscus is receding (vapour-invading-liquid). We may therefore write:
\renewcommand*{\arraystretch}{1.3}
\abceqn{E:Model:theta_+_conditions}{
\left\{
\begin{matrix}
\theta_\pm = \theta_a & \pm \dot{x}_\pm > 0,\\
\theta_r \leq \theta_\pm \leq \theta_a  &\dot{x}_\pm =  0,\\
\theta_\pm = \theta_r & \pm\dot{x}_\pm < 0,
\end{matrix} \right.} where we have accounted for the inherent asymmetry that the meniscus at $x=x_\pm$ is advancing (receding) when $\pm \dot{x}_\pm>0$ ($<0$). \ab{The assumption that the advancing and receding angles are independent of speed is consistent with experimental observations that, at least for moderate capillary numbers, any dependence on meniscus velocity is rather weak \cite{Blake1969, Petrov1991, Hayes1993,Tavana2006CollSuff,Guan2016, Shi2018CollSuff}. An example of a fit to prior experimental data of~\eqref{E:Model:theta_+_conditions} is shown in figure~\ref{fig:hysteresis_graph}.}

%\ab{While this is an extremely simple model, it is consistent with common molecular kinetic models of contact angle hysteresis remains close to the static advancing or receding angle provided that the capillary number remains small. This model is also consistent with several experimental studies which have shown that the variation with velocity is approximately logarithmic \cite{Blake1969,Guan2016} and may even plateau for speeds on the order of $10\mathrm{~\mu m/s}$ \cite{Petrov1991, Hayes1993}. Figure~\ref{fig:hysteresis_graph}a contains a plot of of~\eqref{E:Model:theta_+_conditions} overlain with the experimental data from~\cite{Petrov1991}, demonstrating the agreement between the model~\eqref{E:Model:theta_+_conditions} and experiment in one particular example.}

While this is an extremely simple model, it is consistent with several experimental studies, which have shown that the variation with meniscus velocity is approximately logarithmic \cite{Blake1969,Guan2016}.
 
While the range of values that can be adopted by the contact angle for a stationary meniscus may appear to be ill-constrained by~\eqref{E:Model:theta_+_conditions}, in such situations the contact angle is determined from the pressure, via \eqref{E:Model:LaplacePressure}: the pressure takes the (unique) value that ensures that the pressure gradient (and thus velocity) are zero at the meniscus, thereby determining $\theta_\pm$. We note also that although the conditions for the left and right menisci given in~\eqref{E:Model:theta_+_conditions} may hold independently of one another (giving nine possible cases in total), they must be compatible with conservation of mass; for example,~\eqref{E:Model:theta_+_conditions}c  corresponds to both menisci receding and so is incompatible with channel walls that are deflected inwards.

In our system, the values of the contact angles themselves do not appear;  rather it is $\cos\theta_\pm$ that appears in, for example, \ab{the pressure condition} \eqref{E:Model:LaplacePressure}. For notational convenience, we shall therefore introduce the parameter 
\begin{equation}
\hyspar = \frac{\cos \theta_-}{\cos \theta_+}-1
\end{equation}
as \ab{a} measure of the instantaneous contact angle asymmetry. The maximum value of this parameter is attained with $\theta_+=\theta_a$, $\theta_-=\theta_r$, and so we let
\begin{equation}\label{E:Model:hysteresis_measures}
\hysparmax = \frac{\cos \theta_r}{\cos \theta_a}-1
\end{equation}
be a measure of the asymmetry between the advancing and receding angles. While this measure of contact angle hysteresis is different from the more common definition $\Delta\theta=\theta_a - \theta_r$, the two are closely related: Figure~\ref{fig:hysteresis_graph}b shows that $\hysparmax$ is monotonic increasing in $\Delta\theta$ and $\hysparmax = 0$ if and only if $\theta_a - \theta_r = 0$. \ab{Moreover, for small differences between advancing and receding contact angles, equation~\eqref{E:Model:hysteresis_measures} can be expanded to show that $\hysparmax \propto \Delta\theta$, approximately.}
%\about{Moreover,~\citet{deGennes2004} suggest that surfaces with $\theta_a - \theta_r < 5\si{\degree}$ are relatively clean (shaded in Figure~\ref{fig:hysteresis_graph}); in that case $\hysparmax \ll 1$ and so $\hysparmax\propto \Delta\theta$ (approximately). }

 Note that the instantaneous asymmetry parameter $\hyspar$ has the following properties:  (i) $0\leq \hyspar \leq \hysparmax$ (we will, therefore, often refer to $\hysparmax$ as the maximum contact angle asymmetry) (ii) $\hyspar = 0$ corresponds to $\theta_+ = \theta_-$ (equal contact angles at both menisci) and (iii) $\hyspar = \hysparmax$ if and only if $\theta_+ = \theta_a$ and $\theta_- = \theta_r$ (as we expect for a droplet moving towards the free end of the channel with `$+$' meniscus advancing and `$-$' meniscus receding).

\subsection{Initial Conditions}\label{S:Model:InitialConditions}
%state initial conditions and "we postpone any discussion of the initial conditions on the contact angles 
%make a note that with straight beams initially, there will be an early squeezing regime in which the beams bend to accomodate droplet torque
The problem, which consists of the PDE~\eqref{E:Model:CombinedEq1}--\eqref{E:Model:CombinedEq3} with boundary conditions~\eqref{E:Model:PressureExplicit}--\eqref{E:Model:FreeBC}, kinematic conditions~\eqref{E:Model:Kinematic}, and contact angle condition~\eqref{E:Model:theta_+_conditions} for $h(x,t)$, $\theta_\pm(t)$, $x_{\pm}(t)$, is closed by specifying initial conditions. We assume that the channel is initially undeformed
\begin{equation}\label{E:Model:IC_BeamShape}
h(x,0) = H,
\end{equation}
and the menisci are at known locations
\begin{equation}\label{E:Model:IC_menisci}
x_{\pm}(0) = x_{\pm}^0,
\end{equation} 
which must satisfy the volume constraint
\begin{equation}
2H\left(\xleft^0 - \xright^0\right) = \Omega,
\end{equation} for  \ab{a given} (two-dimensional) droplet volume  $\Omega$.

Note that an initially undeformed channel shape~\eqref{E:Model:IC_BeamShape} provides no torque, whilst the droplet applies a finite torque associated with a non-zero droplet pressure; at early times, the two torques applied to the channel walls --- droplet pressure and restorative from bending --- will not be in balance. We therefore anticipate an early period during which the channel walls respond quickly to this imbalance by bending inwards. During this period, the droplet must spread, with menisci moving in opposite directions; to be consistent with this, we take initially advancing contact angles at the menisci,
\begin{equation}\label{E:Model:IC_contact_angles}
\theta_{\pm}(0) = \theta_a.
\end{equation}
As we shall see, the initial conditions~\eqref{E:Model:IC_contact_angles} result in a scenario in which the contact angle at the `$+$' meniscus is always the advancing angle; accordingly, the  conditions~\eqref{E:Model:theta_+_conditions}b,c for the `$+$' meniscus  are superfluous, but we retain them for completeness (in particular, for a non-wetting droplet, $\theta_{\pm} > 90\si{\degree}$, they must be included in the model).

\subsection{Non-dimensionalization} 
To non-dimensionalize the problem, we use longitudinal and transverse scales based on the channel length $L$ and width $H$, respectively. We use the pressure scale $BH/L^4$ (the characteristic pressure required to bend the channel  wall a distance comparable to the channel width) and the capillary time \ab{scale} $\tau_c = \mu L^2 /(|\gamma \cos \theta_a |H)$ (the characteristic time for liquid of viscosity $\mu$, surface tension $\gamma$ with contact angle $\theta_a$ to imbibe a distance $L$ in a capillary tube of width $H$). We therefore  introduce the dimensionless variables
\begin{equation}\label{E:Model:NonDim:Scalings}
\hat{x}= \frac{1}{L}x, \quad \hat{x}_{\pm} = \frac{1}{L}x_{\pm}, \quad \hat{h} = \frac{1}{H}h, \quad \hat{t} = \frac{1}{\tau_c}t, \quad \hat{p} = \frac{L^4}{B H}p.
\end{equation}

In terms of these dimensionless variables, the system of PDEs~\eqref{E:Model:CombinedEq1}--\eqref{E:Model:CombinedEq3} reads
\begin{align}
0 &= \ddp{^4\hat{h}}{\hat{x}^4} & &0 < \hat{x} < \hat{x}_{-}(\hat{t}),\label{E:Model:NonDim:CombinedEq1} \\
\ddp{\hat{h}}{\hat{t}} &=  \frac{1}{3|\nu|}\ddp{}{\hat{x}}\left(\hat{h}^3 \ddp{^5 \hat{h}}{\hat{x}^5}\right)   & &\hat{x}_{-}(\hat{t}) < \hat{x} <  \hat{x}_{+}(\hat{t}),\label{E:Model:NonDim:CombinedEq2}\\
\qquad \qquad0 &= \ddp{^4\hat{h}}{\hat{x}^4} &  &  \hat{x}_{+}(\hat{t}) < \hat{x} < 1.\label{E:Model:NonDim:CombinedEq3}
\end{align}
Here 
\begin{equation}\label{E:Model:NonDim:nuDefn}
\nu = \frac{\gamma \cos \theta_a L^4}{B H^2}
\end{equation}
is the channel `bendability', and characterizes the ability of the typical capillary pressure within the droplet to bend the channel walls.

In terms of the dimensionless variables, the kinematic conditions~\eqref{E:Model:Kinematic}  read
\begin{equation}\label{E:Model:NonDim:Kinematic}
\dd{\hat{x}_{\pm}}{\hat{t}}  = \left.\frac{1}{3|\nu|}\ddp{^5\hat{h}}{\hat{x}^5} \right|_{\hat{x} = \hat{x}_{\pm}}.
\end{equation}
The channel boundary conditions~\eqref{E:Model:ContinuityBC}--\eqref{E:Model:FreeBC} read
\begin{align}
\hat{h} &= 1,\quad \ddp{\hat{h}}{\hat{x}} = 0, & & \text{at}~ \hat{x} = 0,\label{E:Model:NonDim:BCClamped}\\
\ddp{^2 \hat{h}}{\hat{x}^2} &= 0, \quad\ddp{^3\hat{h}}{\hat{x}^3} = 0, & &   \text{at}~ \hat{x} = 1,\label{E:Model:BCFree}
\end{align}
\abcdeqn{E:Model:NonDim:BCContinuity}{
\left[\hat{h}\right]_{\hat{x}_{\pm}} = \left[\ddp{\hat{h}}{\hat{x}}\right]_{\hat{x}_{\pm}} = \left[\ddp{^2 \hat{h}}{\hat{x}^2}\right]_{\hat{x}_{\pm}} = \left[\ddp{^3 \hat{h}}{\hat{x}^3 }\right]_{\hat{x}_{\pm}} = 0,}
and the pressure boundary condition~\eqref{E:Model:LaplacePressure} reads
\begin{equation}\label{E:Model:NonDim:BCPressure}
\left.\ddp{^4 \hat{h}}{\hat{x}^4}\right|_{\hat{x} = \hat{x}_{\pm}} = -\frac{\nu}{\hat{h}(\hat{x}_{\pm},t)}\frac{\cos \theta_{\pm}}{\cos \theta_a}
\end{equation}

Finally, the dimensionless initial conditions are
\begin{equation}\label{E:Model:NonDim:IC}
\hat{h}(\hat{x}, 0) = 1, \qquad \hat{x}_{\pm}(0) =\hat{x}_{\pm}^0 = \frac{x_{\pm}^0}{L}, \quad \theta_{\pm}(0) = \theta_a.
\end{equation}

The problem~\eqref{E:Model:NonDim:CombinedEq1}--\eqref{E:Model:NonDim:IC} together with the contact angle conditions~\eqref{E:Model:theta_+_conditions} contains five dimensionless parameters: $\nu$, $\hat{x}_+^0$, $\hat{x}_-^0$, $\theta_a$, and $\theta_r$. However, given that in an experiment it is the droplet volume that is specified, it is more natural to use the dimensionless droplet volume 
\begin{equation}
V = \frac{\Omega}{2HL} = \frac{\xright^0 - \xleft^0}{L} = \hat{x}_+^0 - \hat{x}_-^0
\end{equation} to replace  one of the initial meniscus positions. Similarly, it is helpful to use the maximum contact angle asymmetry, $\hysparmax$, in place of one of the contact angles. We therefore consider $(\nu,\hat{x}_+^0,V, \hysparmax, \theta_r)$ to be the pertinent set of dimensionless variables describing a particular experiment. For simplicity, we also set $\theta_r = 0\si{\degree}$  henceforth, since we are concerned with how the presence of contact angle hysteresis (rather than absolute value of the contact angles) affects droplet mobility.

Henceforth, hats are dropped (including on the dimensionless parameters $\hat{x}_-^0$ and $\hat{x}_+^0$) and all variables are assumed to be dimensionless, unless otherwise stated.

\section{Numerical Solutions}\label{S:Numerics}
In this section, we present numerical solutions of the model equations~\eqref{E:Model:NonDim:CombinedEq1}--\eqref{E:Model:NonDim:IC} with contact angle conditions~\eqref{E:Model:theta_+_conditions}. As well as demonstrating how the transitions between the various contact angle conditions should occur in practice, these numerical solutions confirm that our simple model of contact angle hysteresis is sufficient to describe droplet trapping and offers qualitative insight into when this phenomenon occurs.

\begin{figure}[t]
\includegraphics[width = 0.8\textwidth]{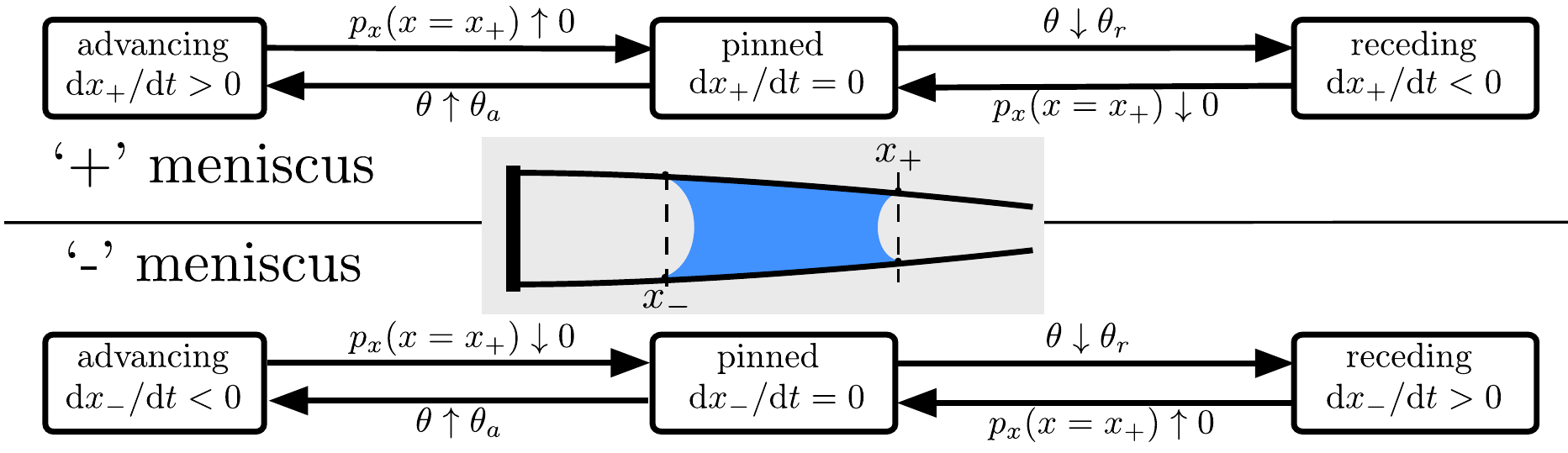}
\caption{Flowcharts of the dynamic events that result in a change of boundary conditions at the `$+$' meniscus (top row) and at the `$-$' meniscus (bottom row) in the model equations describing droplet motion by \bendotaxis~with contact angle hysteresis. The arrow labels indicate the function that triggers the transition; up (down) arrows indicate that the function must increase (decrease, respectively) through the corresponding threshold value. (Recall that the pressure gradient at a meniscus  has an opposite sign to the direction of motion of that meniscus, see~\eqref{E:Model:NonDim:Kinematic}, so sign changes in pressure gradient result in a change in the opposite sense to the meniscus velocity.)  }\label{fig:Numerics:Flowchart}
\end{figure}

The numerical scheme employed here is very similar to that described in the supplementary information of ref.~\cite{Bradley2019PRL}, with transitions  between advancing, pinned, and receding conditions at each meniscus determined by evaluating appropriate event-detection functions at each time-step, as outlined in the flowchart in Figure~\ref{fig:Numerics:Flowchart}. Briefly, the problem is solved numerically by  first transforming it onto one defined only on the droplet region $\xleft(t) < x <\xright(t)$. (This is possible because the shape in the dry regions, $0 < x < \xleft$ and $\xright < x < 1$, can be found analytically and used to give explicit, effective boundary conditions at the menisci that encode the behaviour of the adjacent dry regions.) The resulting `drop-only' problem is then transformed into a flux-conservative form on a time-independent domain by a suitable (time-dependent) rescaling. The resulting partial differential equation is solved numerically with the method of lines~\citep{Schiesser1991}: it is discretized in space, and the resulting set of ordinary differential equations are solved numerically using the \texttt{ode15s} routine implemented in \textsc{MATLAB}. The code used to solve these equations numerically can \ab{be} found at reference~\cite{HysteresisRepo}.

\subsection{Hysteresis dependence}\label{S:Numerics:HysteresisDependence}
%Note to self: have removed the section notation. I think it's unnecessary: we're only going to focus on getting trapped with $\xleft$ pinned (rather than considering getting stuck whilst translating)
\begin{figure}[t]
\centering
\includegraphics[width =0.95\textwidth]{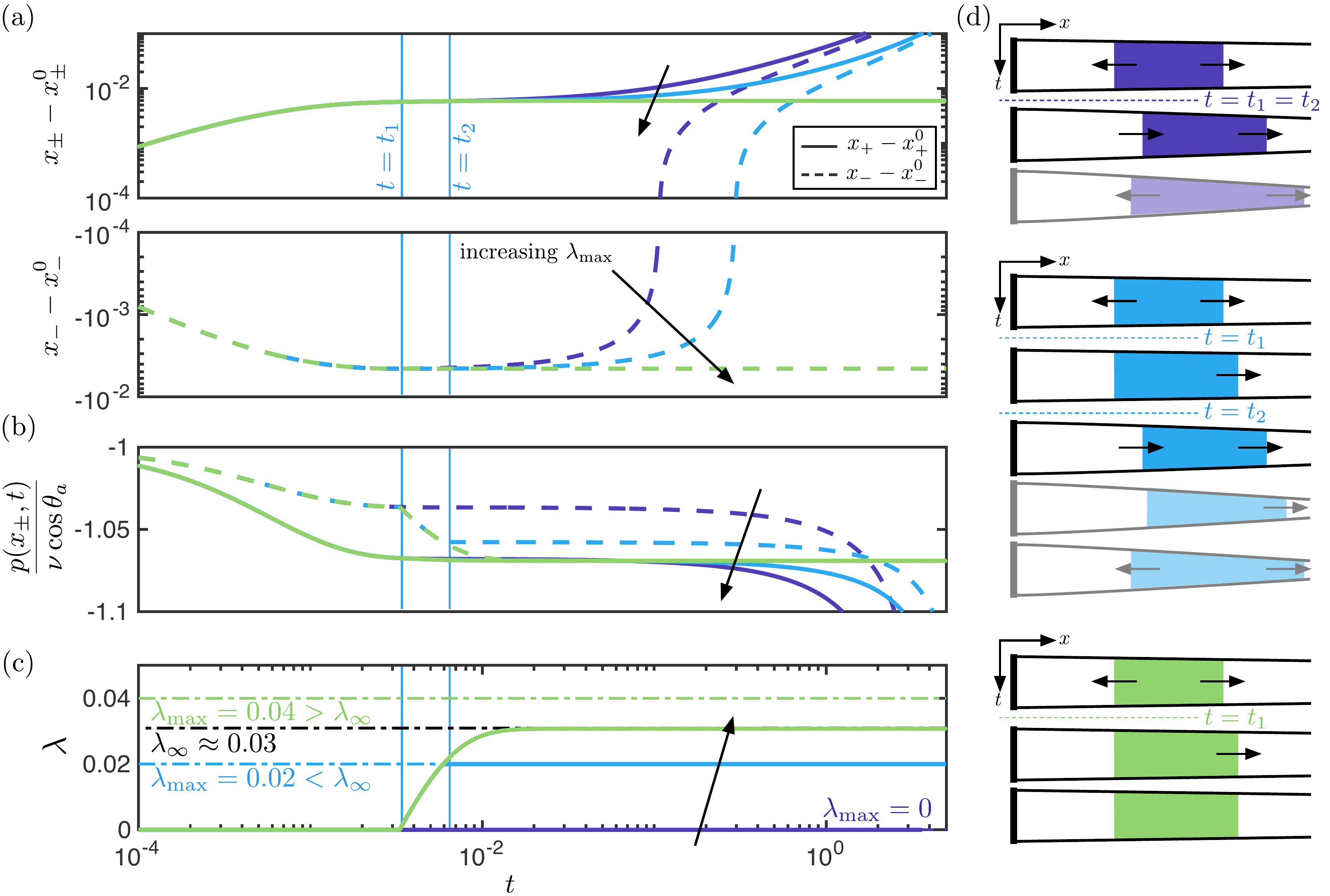}
\caption{An example of the effect of contact angle hysteresis on \bendotaxis~as predicted by our model. The plots in (a)--(c) show the evolution of (a) the displacement of the menisci from their initial positions, (b) the normalized meniscus pressure, and (c) the contact angle asymmetry $\hyspar$.  \ab{(Note that the dashed curves are continuous across the two panels of (a), which use logarithmic axes to facilitate distinction between the curves.)} These predictions are obtained by solving model equations~\eqref{E:Model:NonDim:CombinedEq1}--\eqref{E:Model:NonDim:IC} with $\nu =  4, \xright^0 = 0.65, V = 0.2$ (i.e.~$\xleft^0= 0.45$).  In each plot, solid curves correspond to results for the `$+$' meniscus while dashed curves correspond to  results for the `$-$' meniscus, as indicated by the legend in (a).  Solutions are shown for three different values of the maximum contact angle asymmetry $\hyspar_{\text{max}}$ as follows: $\hyspar_{\text{max}} = 0$ (purple curves, i.e.~no contact angle hysteresis, corresponding to $\theta_a = 0\si{\degree}$), $\hyspar_{\text{max}} = 0.02$ (blue curves, relatively small contact angle hysteresis,  $\theta_a = 10\si{\degree}$), and $\hyspar_{\text{max}} = 0.04$ (green curves, relatively large contact angle hysteresis, $\theta_a = 16\si{\degree}$). \ab{The direction of increasing $\hysparmax$ is indicated by the arrows in each plot.} A log scale on the $y$-axis is used in (a) to aid distinction between the curves. In (c), the coloured horizontal \ab{dot-}dashed lines indicate the corresponding value of $\hyspar = \hysparmax$ and the black dashed line indicates $\hyspar = \hyspar_{\infty}$ (for sufficiently large $\hysparmax$, the contact angle asymmetry reaches $\hyspari$, but evolves no further). The solid blue vertical lines indicate $t = t_1$, the time at which the `$-$' meniscus first becomes pinned, and $t = t_2$, the time at which the `$-$' meniscus first de-pins, for $\hysparmax = 0.02$. The panels in (d) indicate the droplet-channel system schematically throughout the motion, with corresponding droplet colours, as well as the corresponding times $t = t_1, t_2$, where appropriate. Translucent schematics correspond to the second pinning period, which is unimportant for droplet trapping (see main text).}\label{fig:Numerics:ExampleTraces}
\end{figure}

Figure~\ref{fig:Numerics:ExampleTraces} shows the evolution of the position of the menisci, the normalized meniscus pressure, the contact angle asymmetry\ab{,} and the ratio of the channel widths at the menisci as determined from the numerical solution of the model \ab{equations} for three different values of the maximum allowed contact angle asymmetry, $\hysparmax$ (no asymmetry, a relatively small amount, and a relatively large amount -- we shall quantify in due course what small and large means). In each case, identical initial conditions ($\xright^0 = 0.65, \xleft^0 =0.45$) are used.

In the early stages of the motion, the channel walls move inwards in response to the negative capillary pressure; this squeezes the droplet so that both menisci advance (i.e.~move in opposite directions). As a result, $\hyspar = 0$ and solutions with different $\hysparmax$ are identical at early times: the droplet does not have any information about the maximum possible contact angle asymmetry during the early squeezing phase. As the channel continues to deform inwards, the pressure gradient at $\xleft$ decreases, eventually reaching zero so that this meniscus stops moving: the advancing boundary condition~\eqref{E:Model:theta_+_conditions}a is replaced by the pinned boundary condition~\eqref{E:Model:theta_+_conditions}b. At this point, the behaviour of the solutions for different values of $\hysparmax$ diverges.

If $\hysparmax = 0$, the meniscus $\xleft$ is only instantaneously pinned: it immediately turns and moves towards the free end (purple traces in Figure~\ref{fig:Numerics:ExampleTraces}); this scenario is precisely that considered by~\cite{Bradley2019PRL}: the droplet moves along the channel, with both menisci travelling in the same direction (it `translates'), and ultimately reaches the free end. Both menisci increase their speed during this motion, this acceleration occurs despite the low Reynolds number of the motion, being driven by an increasing ratio between the channel widths at the menisci --- the channel is effectively softer at the meniscus closer to the free end ($\xright$) and thus deformations are easier to achieve there (Figure~\ref{fig:Numerics:ExampleTraces}d). As the droplet approaches the free end\ab{,} the meniscus $\xleft$ may be forced, by conservation of mass, to change direction and move once again towards the clamped end, as indicated schematically in Figure~\ref{fig:Numerics:ExampleTraces}d. In Appendix~\ref{A:PostTranslatingDynamics}, we describe these dynamics in more detail and show that this \ab{final} period is not important for droplet trapping, and is thus ignored henceforth.

When there is some contact angle hysteresis, i.e.~$\hyspar_{\text{max}} > 0$, the `$-$' meniscus  remains pinned for a period of time. There are two possible fates for the system beyond this point:  if $\hyspar_{\text{max}}$ is large enough, the meniscus remains pinned for all time and the droplet becomes trapped, whereas for smaller $\hyspar_{\text{max}}$ the meniscus becomes unpinned at some time and the droplet will escape.

In more detail, after the `$-$' meniscus become pinned, the `$+$' meniscus continues to advance and the channel deformation continues to increase, thus reducing the pressure at $\xright$ (increasing the suction) and maintaining $\theta_+ = \theta_a$. To maintain a pinned condition at $\xleft$, the contact angle asymmetry $\hyspar$  increases (the contact angle $\theta_-$ decreases, which acts to increase the magnitude of the suction pressure via the Laplace pressure condition~\eqref{E:Model:NonDim:BCPressure}).  If $\xleft$ remains pinned, the system tends towards an equilibrium, and the contact angle asymmetry tends to a constant value $\hyspari$ (the green curves in Figure~\ref{fig:Numerics:ExampleTraces}).   The value of $\hyspari$ depends on $\nu$, $V$, and $\xright^0$ and emerges from the dynamic model --- it is not possible to determine it a priori; our simulations give $\hyspari \approx 0.03$ for the values $\nu = 4, V = 0.2, \xright^0 = 0.65$ used here.  If $\hysparmax<\hyspari$ however, the system cannot reach this equilibrium and the $\xleft$ meniscus instead de-pins when $\hyspar$ reaches $\hysparmax$ (the blue curves in Figure~\ref{fig:Numerics:ExampleTraces}).  Thereafter, we have $\theta_- = \theta_r$ (while $\theta_+ = \theta_a$ still) and the droplet then accelerates towards the free end of the channel.  (As in the case when $\hysparmax = 0$, there may be a final squeezing phase in which $\xleft$ is forced to reverse direction, but this does not prevent $\xright$ reaching the free end and is not discussed further.)

In summary, when contact angle hysteresis is relatively small, $\hysparmax < \hyspari$ the droplet ultimately escapes the channel by translating to the free end. Otherwise, the maximum allowed asymmetry is relatively large, $\hysparmax > \hyspari$, then the droplet will be trapped indefinitely: it will remain part-way along the channel. It is also interesting to note that the simulation with $\hysparmax = 0.02$ takes approximately twice as long to reach the free end compared to the simulation with no hysteresis, $\hysparmax = 0$; this suggests \ab{that droplet dynamics have} strong sensitivity to contact angle hysteresis even when droplets ultimately escape; \ab{the slowing down results from the fact that the difference in Laplace pressure between the mensici is reduced in the case with contact angle hysteresis (compared to that without) and hence can only permit a slower flow}. Although we are primarily concerned with droplet trapping in this paper, we note that this finding suggests that experimental studies of the dynamics of self-propelled droplets must be careful to minimize contact angle hysteresis if its effect is to be neglected in the corresponding models. Reference~\citep{Bradley2019PRL} reported droplet speeds that were systematically lower than model predictions; the results presented here suggest that a moderate contact angle hysteresis may be responsible for this discrepancy. %do we want actual numbers - no! We didn't actually measure the hysteresis in the expts

\subsection{Initial position dependence}
\begin{figure}[t]
\includegraphics[width = 0.95\textwidth]{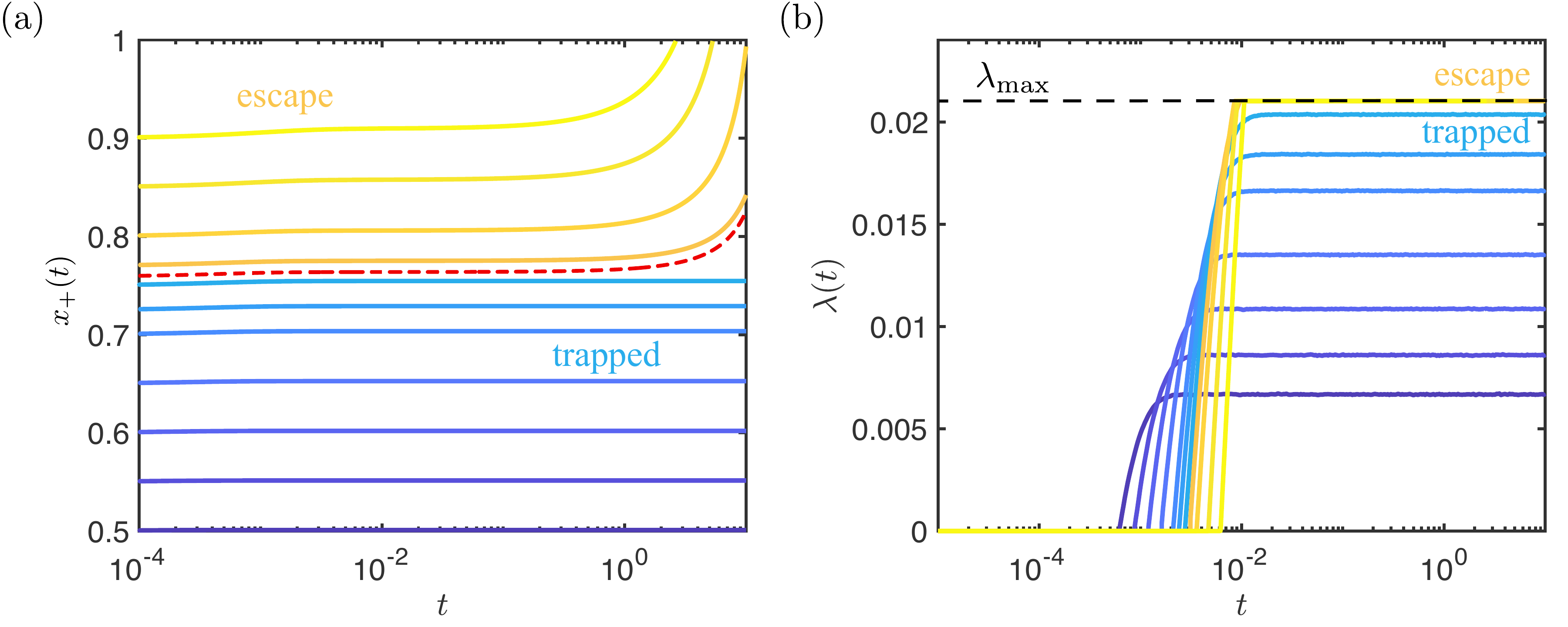}
\caption{Temporal evolution of the (a) meniscus position $\xright(t)$ and (b) contact angle asymmetry $\hyspar$ determined from numerical solution of our model equations~\eqref{E:Model:NonDim:CombinedEq1}--\eqref{E:Model:NonDim:IC},
with $V = 0.2$, $\nu = 2$ and $\hyspar_{\text{max}} \approx 0.02$; results with different initial meniscus positions in the range $0.5\leq \xright^0 \leq0.9$ are shown.  Note that droplets starting close to the free end ($\xright^0$ sufficiently large, yellow hue curves) escape, whilst those starting closer to the base ($\xright^0$ sufficiently small, blue hue curves) are trapped indefinitely. The red dashed trajectory corresponds approximately to $\xright^0 = \xright^{0,\text{escape}}$, the smallest value of $\xright^0$ at which the droplet escapes (see \S\ref{S:MobileDroplets}).}\label{fig:Numerics:initialpos}
\end{figure}
 
To illustrate the effect of the initial droplet position on its ultimate fate, Figure~\ref{fig:Numerics:initialpos} shows the  numerically obtained droplet trajectories, $\xright(t)$, together with the corresponding evolution of the contact angle asymmetry, $\hyspar(t)$, for various initial meniscus positions in the range $0.5 \leq \xright^0 \leq 0.9$. 
As observed previously, all droplets undergo an initial squeezing phase during which both menisci advance and $\lambda=0$. Once $\lambda$ starts to increase, however, the effect of the initial position becomes apparent:  for droplets that start sufficiently close to the free end, $\hyspar$ reaches $\hysparmax$, at which point  the droplet begins to translate and ultimately escapes. In contrast, for droplets that start closer to the base (smaller values of $\xright^0$), $\hyspar$ reaches $\hyspari$ before $\hysparmax$ and the droplet is trapped. Note that here, as before, droplets that are trapped remain close to their initial positions indefinitely.
 
This figure suggests that the final `trapped' value of the contact angle asymmetry, $\hyspari$, is an increasing function of $\xright^0$, as we might expect: a greater contact angle difference will be needed to maintain the pinned state when the droplet begins nearer the free end of the channel, which is `softer' than the clamped end. The deformation in the pinned state is also an increasing function of bendability $\nu$ and volume $V$ (qualitatively, larger $\nu$ means a stronger pull on the beams, while larger $V$ increases the area over which this pull is applied). Accordingly, the effect of changes in $\nu$ and $V$ on the ultimate fate of the droplet is similar to that of the initial droplet position: for given values of $\xright^0$ and $\hysparmax$, droplets of sufficiently large volume or in systems with sufficiently large bendability will escape, whilst others will not; in other words,   $\hyspari$ is an increasing function of $V$ and $\nu$ (data not shown).

\subsection{Discussion}

The results shown in this section confirm our intuition that when hysteresis is sufficiently strong, droplets may get trapped part way along the channel.  The numerical solutions of our model highlight three important features of the trapping mechanism that appear to be generic:
firstly, the system always passes through a squeezing period during which both menisci advance until $\xleft$ is pinned; secondly, there is a contact angle asymmetry, $\hyspari$, required to maintain the meniscus at $\xleft$ in a pinned condition  indefinitely; and, thirdly, if $\hysparmax \leq \hyspari$, the maximum contact angle asymmetry is not enough to pin the droplet indefinitely and so the droplet begins to translate with $\xright$ advancing, ultimately reaching the free end (the droplet escapes). (Equivalently, if $\hysparmax > \hyspari$ the droplet remains in the pinned state and the droplet is trapped.) Determining the value of $\hyspari$ is therefore critical to answering the central question of this paper: in which regions of $(\nu, V, \xright^0, \hysparmax)$ parameter space do droplets get trapped within the channel as a result of contact angle hysteresis? While the value of $\hyspari$ cannot be determined \emph{a priori}, but  emerges as part of the solution, we can approximate it by exploiting the observation that trapped droplets do not move significantly from their initial positions. Before we are able to do so, however, we must consider the configurations occupied when droplets are trapped, i.e.~the equilibria of the system; we turn to this now.

\section{Equilibrium configurations}\label{S:Equilibria}
The numerical solutions presented in \S\ref{S:Numerics} suggest that droplets can be trapped indefinitely if the contact angle hysteresis is sufficiently large, or if droplets start sufficiently close to the clamped end. In this section, we consider these trapped equilibrium states. We aim to determine when equilibria exist and analyze their linear stability, with a view to (i) verifying that the numerical solutions presented in \S\ref{S:Numerics} are indeed converging to true equilibria (rather than simply slowly evolving transients) and (ii) determining the linear stability of these equilibria.

In this section, we consider equilibrium configurations with contact angle conditions reflecting those observed in the motion immediately preceding droplet trapping, i.e. we assume that $\thetaright =\theta_a$ (advancing) and $\theta_r < \thetaleft< \theta_a$ (pinned). We denote the contact angle asymmetry that this encodes by $\hyspar = \hyspare$; the results of this section are then expected to be pertinent provided that $\hyspare$ is attainable, i.e.~provided that $\hyspare \leq \hysparmax$. Note that we use $\hysparmax$ to determine the equilibrium states recorded by the time-dependent solution in the previous section. However, the equilibrium attained emerges dynamically and may correspond to any value up to $\hysparmax$; we analyze equilibria for a given $\hyspare$ in this section and will then observe in \S\ref{S:MobileDroplets} that the value of $\hyspari$ typically corresponds to an equilibrium that is close to the initial condition.

\subsubsection{Equations for equilibrium}
The equations that must be satisfied by equilibrium configurations can be recovered as the steady case of the dynamic problem (equations~\eqref{E:Model:NonDim:CombinedEq1}--\eqref{E:Model:NonDim:BCPressure}). The problem for the equilibrium channel wall shape $h_e(x)$ with menisci located at $x_{\pm} = \xpmeq$ is
\begin{align}
\dd{^4 h_e}{x^4} &= 0 & & 0 < x < \xlefteq,\label{E:Equilibria:Eqs:Beam1}\\
\dd{^4 h_e}{x^4} &= p_0 & & \xlefteq < x < \xrighteq,\label{E:Equilibria:Eqs:Beam2}\\
\dd{^4 h_e}{x^4} &= 0 & &\xrighteq < x < 1,\label{E:Equilibria:Eqs:Beam3}
\end{align}
where $p_0$ is the droplet pressure. This pressure is constant throughout the droplet, and must satisfy
\begin{equation}\label{E:Equilibria:Eqs:BeamP}
p_0 = -\left.\frac{\nu}{h_e}\right|_{x = \xrighteq}  = -\left.\frac{\nu(1 + \hyspare) }{h_e}\right|_{x = \xlefteq}.
\end{equation}

The problem~\eqref{E:Equilibria:Eqs:Beam1}--\eqref{E:Equilibria:Eqs:BeamP} must be solved subject to further boundary conditions
\begin{equation}\label{E:Equilibria:Eqs:ClampedBC}
h_{e} = 1,\quad  \dd{h_e}{x} = 0 \quad \text{at}~x = 0, 
\end{equation}
and
\begin{equation}\label{E:Equilibria:Eqs:FarEndBCFree}
\dd{^2 h_e}{x^2} =\dd{^3 h_e}{x^3}=0 \quad \text{at}~x = 1, 
\end{equation}
with continuity conditions
\begin{equation}\label{E:Equilibria:Eqs:Continuity}
\left[h_e\right]_{\xpmeq} = \left[\dd{h_e}{x}\right]_{\xpmeq} = \left[\dd{^2 h_e}{x^2}\right]_{\xpmeq}=\left[\dd{^3 h_e}{x^3}\right]_{\xpmeq} =0.
\end{equation}

The solution $h_e(x)$ must also satisfy the global volume constraint
\begin{equation}\label{E:Equilibria:Eqs:Volume}
V = \int_{\xlefteq}^{\xrighteq}h_e(x)~\mathrm{d}x,
\end{equation} 
and the beam ends must not touch, 
\begin{equation}\label{E:Equilibria:Eqs:OpenEnds}
h_e(1)>0.
\end{equation}

Note that by re-arranging~\eqref{E:Equilibria:Eqs:BeamP}, the contact angle asymmetry $\hyspare$ can be expressed as a geometric constraint on the solution:
\begin{equation}\label{E:Equilibria:Eqs:hyspar_geometry}
\hyspare = \frac{h_e(\xlefteq)}{h_e(\xrighteq)} - 1.
\end{equation}
The condition~\eqref{E:Equilibria:Eqs:hyspar_geometry} is useful for understanding the equilibrium maps presented in \S\ref{S:Equilibria:EquilibriumMaps}, which indicate the regions of parameter space in which solutions \ab{to~\eqref{E:Equilibria:Eqs:Beam1}--\eqref{E:Equilibria:Eqs:OpenEnds}} exist.

\subsubsection{Equilibria with $0<\xlefteq\ll1$}
The equations for equilibrium~\eqref{E:Equilibria:Eqs:Beam1}--\eqref{E:Equilibria:Eqs:OpenEnds} do not have an analytic solution in general. However, analytic progress can be made if we impose (instead of solving for) $0<\xlefteq\ll1$, which serves as a useful limiting case in the following.

In this case, we must have $h_e(\xlefteq)\approx h_e(0)=1$, using \eqref{E:Equilibria:Eqs:ClampedBC} and so, combining with the pressure condition~\eqref{E:Equilibria:Eqs:BeamP} we find that the equilibrium pressure within the droplet is simply $p_0 =- \nu (1 + \hyspare)$. We can then readily find an  analytic solution for the channel shape,
\begin{equation}\label{E:Equilibria:Eqs:analytic_sol}
h_e(x) = 1-\frac{\nu(\hyspare +1)}{24}\times\begin{cases} (x-\xrighteq)^4 + 4\left(\xrighteq\right)^3(x-\xrighteq) + 3\left(\xrighteq\right)^4 & 0 < x < \xrighteq\ab{,}\\
\left(\xrighteq\right)^3 (4x-\xrighteq) & \xrighteq < x <1,
\end{cases}
\end{equation}
where the meniscus position $\xrighteq$ must ensure that the pressure condition~\eqref{E:Equilibria:Eqs:BeamP} is satisfied, requiring
\begin{equation}\label{E:Equilibria:Eqs:analytic_sol_meniscus_position}
\left.\xrighteq\right.^4 = \frac{8\hyspare}{\nu (1 + \hyspare)^2}.
\end{equation}
To satisfy the volume constraint~\eqref{E:Equilibria:Eqs:Volume} we require that $\hyspare$ satisfies

\begin{equation}\label{E:Equilibria:Eqs:analytic_sol_volume}
\frac{\hyspare}{(1+\hyspare)^2}\left(\frac{3\hyspare + 5}{5\hyspare + 5}\right)^4 = \frac{\nu V^4}{8}.
\end{equation} Note that~\eqref{E:Equilibria:Eqs:analytic_sol_volume} only has a solution for $\nu V^4 \lesssim 1.01$; configurations with $\nu V^4 > 1.01$ violate $\xright < 1$: the deformation they create is too large to accommodate the droplet within the channel.

Since we consider $\lambda_e$ to be given here, equation~\eqref{E:Equilibria:Eqs:analytic_sol_volume}  can be used to determine the corresponding $\nu$ for a given $V$. With this constraint on $\nu$, the open ends constraint~\eqref{E:Equilibria:Eqs:OpenEnds} holds provided that
\begin{equation}\label{E:Equilibria:Eqs:analytic_sol_open_ends}
V > \frac{4\hyspare(3\hyspare + 5)}{5(\hyspare+1)(4\hyspare + 3)}.
\end{equation}

We now move on to consider when equilibria with a particular value of $\hyspare$ are possible for different values of the parameters $\nu$ and $V$. This will be guided by the conditions \eqref{E:Equilibria:Eqs:analytic_sol_volume} and \eqref{E:Equilibria:Eqs:analytic_sol_open_ends}. 

\subsubsection{Equilibrium Maps}\label{S:Equilibria:EquilibriumMaps}
\begin{figure}[t]
\centering
\includegraphics[width = 0.9\textwidth]{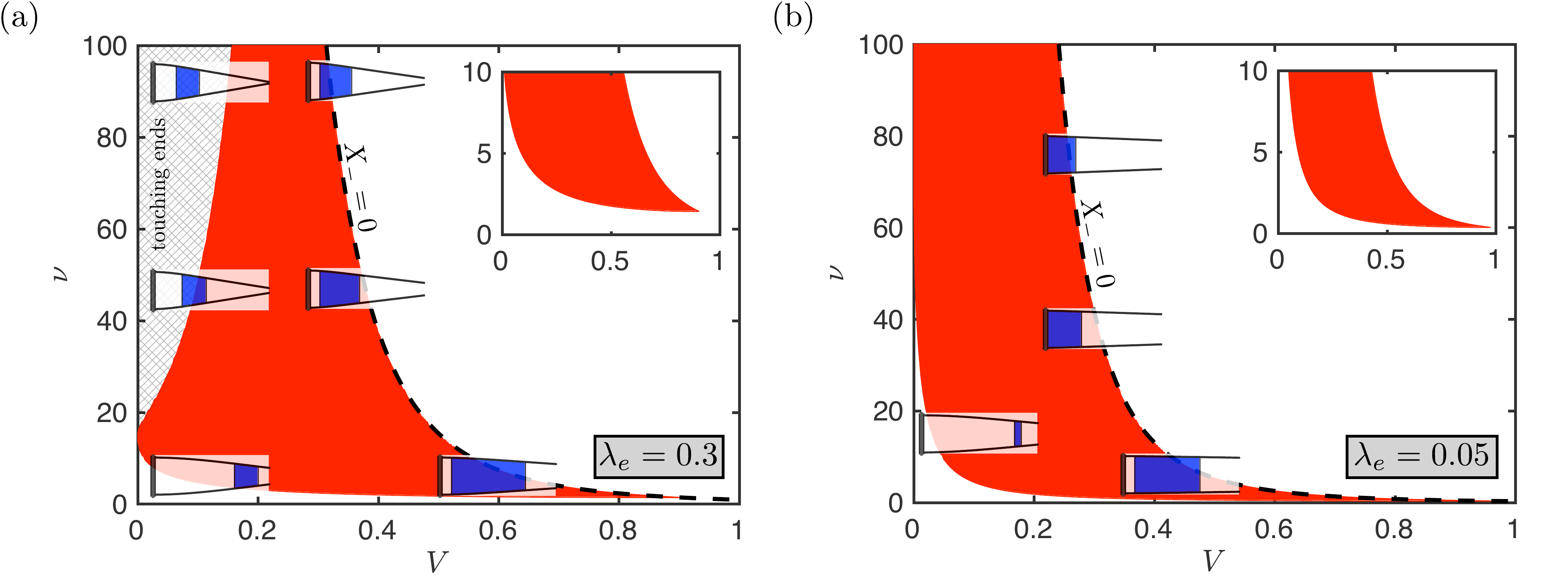}
\caption{Diagrams showing regions of $(V, \nu)$ space for which solutions of equations~\eqref{E:Equilibria:Eqs:Beam1}--\eqref{E:Equilibria:Eqs:Volume}  exist with (a) $\hyspare = 0.3$ (corresponding to an advancing angle $\theta_a \approx 40\si{\degree}$ and (b) $\hyspare = 0.05$ ($\theta_a\approx 18\si{\degree}$). The schematic diagrams indicate the shape of the configuration close to that region of parameter space. The black dashed curve indicates~\eqref{E:Equilibria:Eqs:analytic_sol_volume}, corresponding to equilibria with $X_- = 0$ (equations~\eqref{E:Equilibria:Eqs:analytic_sol_meniscus_position}--\eqref{E:Equilibria:Eqs:analytic_sol_open_ends}). The hatched region in (a) indicates where solutions to~\eqref{E:Equilibria:Eqs:Beam1}--\eqref{E:Equilibria:Eqs:Volume} exist that violate the no touching condition~\eqref{E:Equilibria:Eqs:OpenEnds}. In each plot the inset contains a close-up of the main figure in the region $0 < \nu < 10$, $0<V<1$. }\label{fig:ExampleRegimeDiagram} 
\end{figure}

Equilibrium configurations are obtained numerically. Full details \ab{of this procedure} can be found in Appendix~\ref{A:FindingEquilibria}, but we note that, for convenience, we do not solve the (non-linear) equilibrium equations~\eqref{E:Equilibria:Eqs:Beam1}--\eqref{E:Equilibria:Eqs:Volume} for given $(\nu, V, \hyspare)$ directly; rather we specify one of the meniscus positions (typically $\xlefteq$), and then solve equations~\eqref{E:Equilibria:Eqs:Beam1}--\eqref{E:Equilibria:Eqs:Continuity}; the volume associated with each equilibrium is then readily calculated using~\eqref{E:Equilibria:Eqs:Volume}, and the equilibrium is retained only if it satisfies the open end condition~\eqref{E:Equilibria:Eqs:OpenEnds}. By sweeping over all permissible values of $\xlefteq$, we pick up all possible solutions of~\eqref{E:Equilibria:Eqs:Beam1}--\eqref{E:Equilibria:Eqs:Volume}. We find that for a given $(\nu, V, \hyspar_e)$ if a solution to~\eqref{E:Equilibria:Eqs:Beam1}--\eqref{E:Equilibria:Eqs:Volume} satisfying~\eqref{E:Equilibria:Eqs:OpenEnds} exists, then that solution is unique.

In Figure~\ref{fig:ExampleRegimeDiagram} we show equilibrium maps that indicate the regions of $(V, \nu)$ space in which equilibria exist\ab{,} for two different values of $\hyspare$, corresponding to very high hysteresis ($\hyspare = 0.3$, Figure~\ref{fig:ExampleRegimeDiagram}a) and a more typical value ($\hyspare = 0.05$, Figure~\ref{fig:ExampleRegimeDiagram}b). For completeness, we present data for $0 < \nu < 100$ but  in practice droplets in channels with $\nu \gtrsim 10$ are prone to trapping themselves by closing the channel walls during the motion~\citep{Bradley2019PRL}; we include as insets in Figure~\ref{fig:ExampleRegimeDiagram} the same equilibrium maps zoomed into the region $0 < \nu < 10$ of parameter space in which configurations are not susceptible to this `geometric trapping', which are of most interest here. 

We can rationalize the shape of these equilibrium maps by considering $\hyspare$ to be a geometric constraint on the capillary induced wall deflections, as encoded by equation~\eqref{E:Equilibria:Eqs:hyspar_geometry}; capillary induced wall deformations, whose size depends on the strength of surface tension (via $\nu$), the length over which the force is applied (via $V$) and the position of the droplet (via $\xrighteq$) must exactly balance the contact angle asymmetry $\hyspare$.  At small $\nu$ (weak surface tension), the Laplace pressure in the droplet is not able to create enough deflection to satisfy the geometric constraint~\eqref{E:Equilibria:Eqs:hyspar_geometry}, regardless of the droplet's size or position in the channel, and so no equilibria exist.  As $\nu$ increases, equilibria first appear with $\xrighteq = 1$ (see schematics in Figure~\ref{fig:ExampleRegimeDiagram}), since droplets are able to create the largest deflection when they are at the free end of the channel. This lower boundary of $\nu$ values is decreasing in $V$ (insets in Figure~\ref{fig:ExampleRegimeDiagram}) because larger droplets can generate the same deflection by applying a lower pressure (smaller $\nu$) over a larger area. Similarly, the minimum value of $\nu$ (for a fixed $V$) at which equilibrium configurations exist is smaller for smaller $\hyspare$ --- less deflection is needed to satisfy the geometric constraint~\eqref{E:Equilibria:Eqs:hyspar_geometry}, which can therefore be achieved with a lower surface tension.

As $\nu$ increases (maintaining a constant volume $V$), equilibrium configurations have droplets closer to the base, where the higher bendability is countered by pressure being applied over relatively stiffer sections of the channel, and the channel width at the free end $x = 1$ is smaller. When $\nu$ is sufficiently large, equilibria fail to exist because either (i) the channel width at the free end reaches zero (the two ends touch, violating the no contact condition~\eqref{E:Equilibria:Eqs:OpenEnds}; visible in Figure~\ref{fig:ExampleRegimeDiagram} only for the larger value of $\hyspare$), or (ii) the lower meniscus reaches the base, $\xlefteq = 0$, so that the droplet can move no further to offset increasing bendability; this is shown by the dashed curve in Figure~\ref{fig:ExampleRegimeDiagram} and is expressed  analytically by \eqref{E:Equilibria:Eqs:analytic_sol_open_ends}. 

\begin{figure}[t]
\centering
\includegraphics[width = 0.9\textwidth]{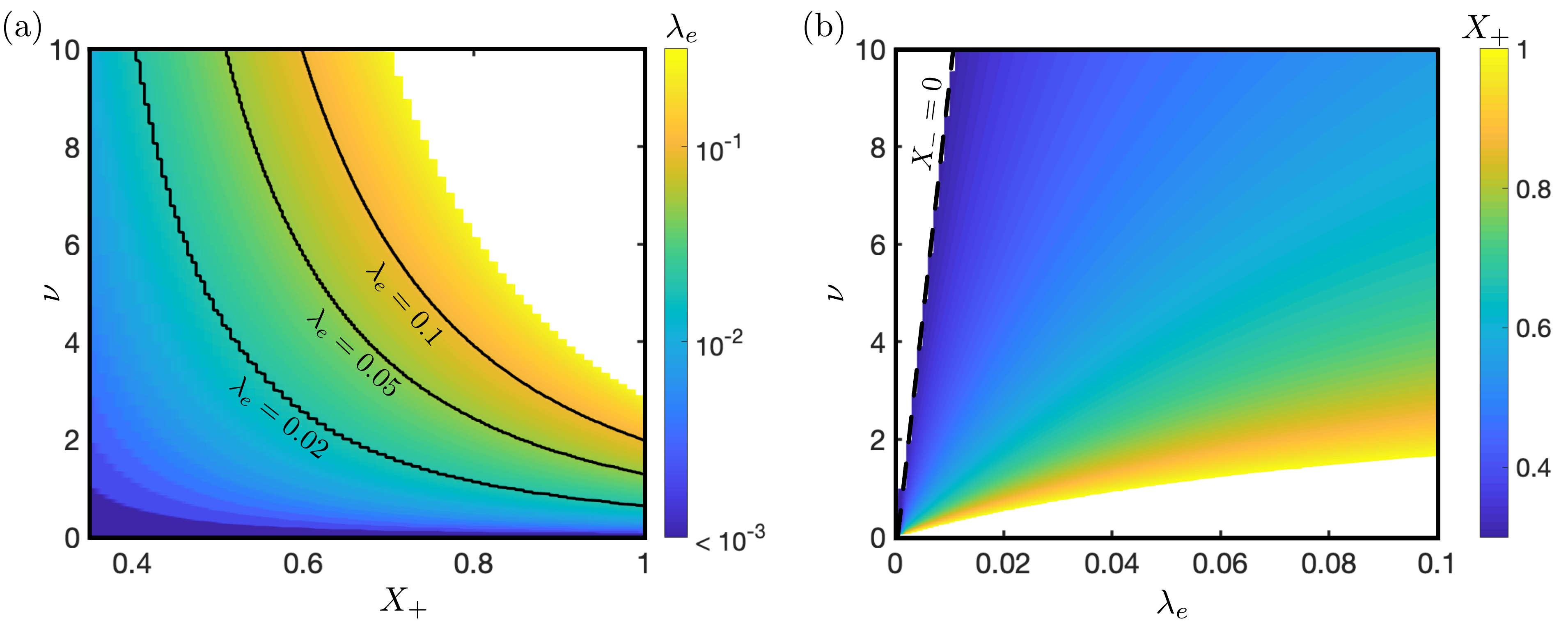}
\caption{Equilibrium maps indicating (a) the value of $\hyspare$ associated with equilibria in $(\xrighteq, \nu)$ space and (b) the value of $\xrighteq$ associated with equilibria in $(\hyspare, \nu)$ space (with $V = 0.3$ in both cases). In  both plots, blank regions indicate areas where equilibria do not exist. The black dashed line in (b) indicates equation~\eqref{E:Equilibria:Eqs:analytic_sol_volume}, corresponding to equilibria with $\xlefteq = 0$.}\label{fig:regime_diagrams_xright_and_hyspar_vs_nu}
\end{figure}

In Figure~\ref{fig:regime_diagrams_xright_and_hyspar_vs_nu} we show two other ways of presenting equilibrium maps. Firstly, in Figure~\ref{fig:regime_diagrams_xright_and_hyspar_vs_nu}a, we plot the value of $\hyspare$ associated with equilibria in $(\xrighteq, \nu)$ space, for the $\mathcal{O}(1)$ values of the bendability $\nu$ that  are of most interest. This plot indicates that equilibria in which the droplet is located closer to the free end are associated with a larger $\hyspare$ (encoding a larger difference between the channel widths at the menisci) and that this difference is more pronounced for larger $\nu$. 

Secondly, in Figure~\ref{fig:regime_diagrams_xright_and_hyspar_vs_nu}(b), we plot the value of $\xrighteq$ associated with equilibria in $(\hyspare, \nu)$ space. In particular, this plot indicates that equilibria do not exist when the contact angle asymmetry $\lambda_e$ is too large (the droplet is not able to create enough deflection to satisfy~\eqref{E:Equilibria:Eqs:hyspar_geometry}, regardless of where it sits in the channel) or too small (the droplet always creates too much deflection, regardless of where it sits in the channel). 

\subsubsection{Stability}
We analyze the linear stability of equilibria by letting \newcommand{\xrighteqpert}{\zeta} %give the perturbation an clearly different name to avoid confusion/annoying superscripts?
\begin{equation}\label{E:Equilibria:Stability:perturbation}
h = h_e(x) + \epsilon e^{\sigma t} h_1(x), \qquad \xright(t) = \xrighteq +  \epsilon  e^{\sigma t} ,
\end{equation} 
where $\epsilon \ll 1$ is arbitrary, in the model equations. 

After a standard linearization procedure, the problem for the wall perturbation, $h_1(x)$, becomes
\begin{align}
0 &= \dd{^4 h_1}{x^4} & & 0 < x < \xlefteq,~\xrighteq < x < 1,\label{E:Equilibria:Stability:PDE1}\\
3\nu\sigma h_1 &=\dd{}{x} \left(h_e^3 \dd{^5 h_1}{x^5}\right) & &\xlefteq < x < \xrighteq,\label{E:Equilibria:Stability:PDE2}
\end{align}
with boundary conditions,
\begin{align}
h_1 & = \dd{h_1}{x} = 0 & &\text{at}~x = 0,\label{E:Equilibria:Stability:clampedBC}\\
\dd{^2 h_1}{x^2} &=\dd{^3 h_1}{x^3} = 0 & &\text{at}~x = 1,\label{E:Equilibria:Stability:freeBC}
\end{align}
and continuity conditions
%\left[h_1 + \xpmeq^1\dd{h_e}{x}\right]_{\xpmeq^-}^{\xpmeq^+} = \left[\dd{h_1}{x} + \xpmeq^1\dd{^2 h_e}{x^2}\right]_{\xpmeq^-}^{\xpmeq^+} &=0, \\ \left[\dd{^2 h_1}{x^2 } +\xpmeq^1 \dd{^3 h_e}{x^3}\right]_{\xpmeq^-}^{\xpmeq^+} = \left[\dd{^3 h_1}{x} +\xpmeq^1 \dd{^4 h_e}{x^4}\right]_{\xpmeq^-}^{\xpmeq^+}  &= 0.\label{E:Equilibria:Stability:continuity2}
\begin{align}
\left[h_1 \right]_{\xlefteq} = \left[\dd{h_1}{x} \right]_{\xlefteq} = \left[\dd{^2 h_1}{x^2} \right]_{\xlefteq}= \left[\dd{^3 h_1}{x^3} \right]_{\xlefteq} = 0,\\
\left[h_1 \right]_{\xrighteq} = \left[\dd{h_1}{x} \right]_{\xrighteq} = \left[\dd{^2 h_1}{x^2} \right]_{\xrighteq} = \left[\dd{^3 h_1}{x^3} + \xrighteqpert \dd{^4 h_e}{x^4} \right]_{\xrighteq} = 0.\label{E:Equilibria:Stability:continuity2}
\end{align}
Here we have made extensive use of the continuity of the equilibrium shape~\eqref{E:Equilibria:Eqs:Continuity}.  The perturbation must conserve volume, so we require
\begin{equation}\label{E:Equilibria:Stability:kinematic}
0 = \int_{\xlefteq}^{\xrighteq}h_1 ~\mathrm{d}x - \xrighteq^1 h_e(\xrighteq)
\end{equation}
The final (pressure) boundary conditions on~\eqref{E:Equilibria:Stability:PDE1}--\eqref{E:Equilibria:Stability:PDE2}, at $x = \xpmeq$, reflect the fact that the meniscus at $\xlefteq$ is pinned, and the meniscus at $\xrighteq$ is free to move:
\begin{align}
\dd{^5h_1}{x^5} &=0 & &\text{at}~x = \xlefteq,\label{E:Equilibria:Stability:bc1}\\
\dd{^4 h_1}{x^4} &= \frac{\nu}{h_e^2} \left(\dd{h_e}{x} + h_1\right)& &\text{at}~x = \xrighteq.\label{E:Equilibria:Stability:bc2}
\end{align}

\begin{figure}
\centering
\includegraphics[width = 0.95\textwidth]{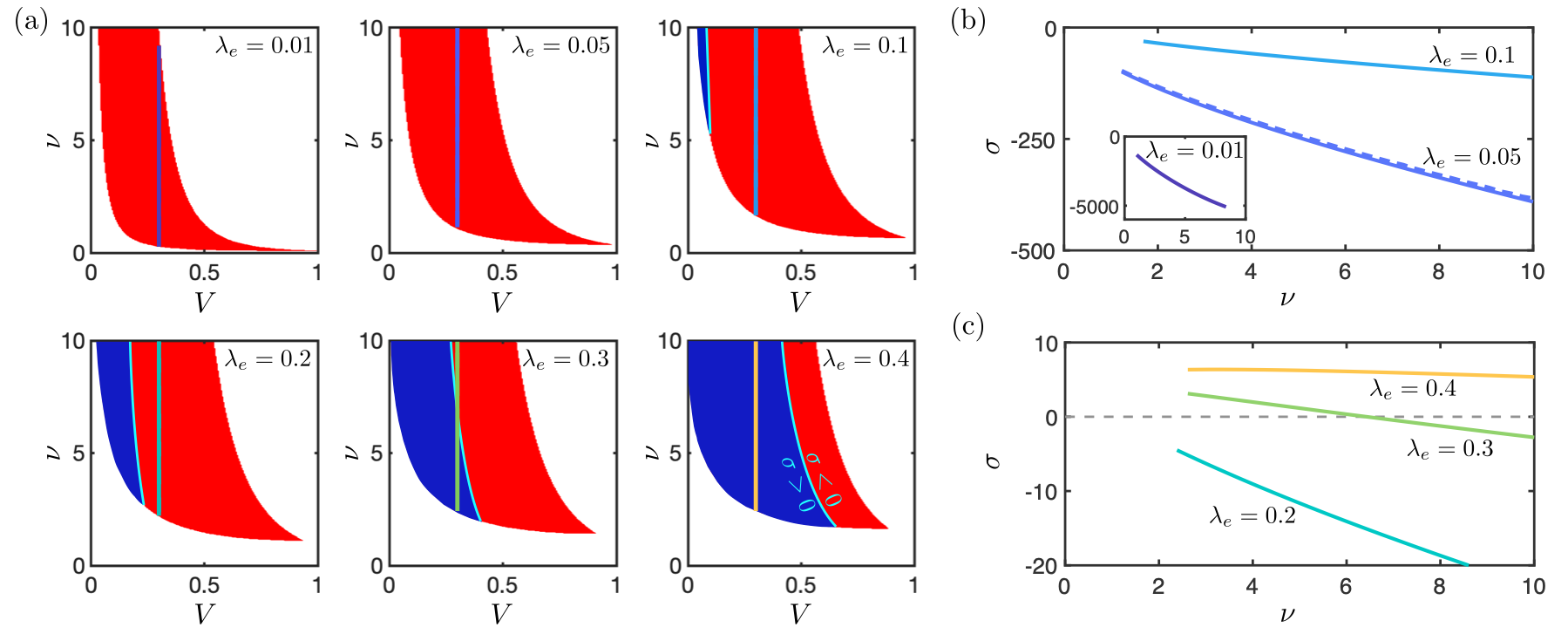}
\caption{(a) Equilibrium maps showing regions of $(V, \nu)$ space for which solutions of~\eqref{E:Equilibria:Stability:PDE1}--\eqref{E:Equilibria:Stability:bc2} exist for $\hyspar = 0.01,~0.05,~0.1,~0.2,~0.3$, and $0.4$, as indicated in the top right of each map. Red regions correspond to linearly stable equilibria, and blue regions correspond to linearly unstable equilibria; the two regions are separated by the cyan curves indicating solutions of~\eqref{E:Equilibria:Stability:solvability}, which correspond to $\sigma = 0$. (b,c and inset) Growth rates $\sigma$, obtained by numerically solving the BVP~\eqref{E:Equilibria:Stability:PDE1}--\eqref{E:Equilibria:Stability:bc2}, at $\nu$ values along the corresponding coloured lines in (a). The blue dashed curve in (b) indicates an estimate of the growth rate of perturbations with $\hyspare=0.05$ obtained by solving the full model equations numerically and performing an exponential fit to the meniscus displacement at early times. }\label{fig:Stability}
\end{figure}

The boundary value problem (BVP) given by~\eqref{E:Equilibria:Stability:PDE1}--\eqref{E:Equilibria:Stability:bc2} must be solved numerically; we use the \texttt{BVP4c} routine implemented in \textsc{matlab}, which returns the growth rate $\sigma$ as part of the solution. Numerical solutions of the BVP agree well (compare the blue solid and dashed curves in Figure~\ref{fig:Stability}b) with numerical solutions of the full model equations, in which the growth rate is determined by an exponential fit to the meniscus trajectory at early times and the perturbation away from equilibrium is applied as a sinusoidal perturbation to the channel shape that preserves volume. Note that in~\eqref{E:Equilibria:Stability:perturbation}, we neglected a variation in the contact angle $\theta_-$; agreement between numerical solutions of the full model equation and the BVP suggest that this variation is not important.

We do not dwell further on solutions of the BVP, however, because we are primarily interested in the stability of equilibria (i.e.~the sign of $\sigma$), rather than the time scale of evolution (the magnitude of $1/\sigma$). It is instructive to consider instead the marginal stability problem given by~\eqref{E:Equilibria:Stability:PDE1}--\eqref{E:Equilibria:Stability:bc2} with $\sigma = 0$. In this case~\eqref{E:Equilibria:Stability:PDE2} can be integrated directly to give
\begin{equation}\label{E:Equilibria:Stability:integrate_once}
h_e^3 \dd{^5 h_1}{x^5} = 0,
\end{equation}
where we have used~\eqref{E:Equilibria:Stability:bc1} to set the constant of integration to zero. From~\eqref{E:Equilibria:Stability:integrate_once} and the remaining boundary conditions (\eqref{E:Equilibria:Stability:clampedBC}--\eqref{E:Equilibria:Stability:continuity2} and~\eqref{E:Equilibria:Stability:bc2}), we can express $h_1$ in terms of $h_e$. The conservation of volume equation~\eqref{E:Equilibria:Stability:kinematic} then becomes a non-linear constraint of the form
\begin{equation}\label{E:Equilibria:Stability:solvability}
S(\nu, V, \hyspare) = 0.
\end{equation}

Numerical solutions of~\eqref{E:Equilibria:Stability:solvability} are shown as cyan curves in the  equilibrium maps shown in Figure~\ref{fig:Stability}a. We see that for small to moderate values of $\hyspare$ there are no solutions of~\eqref{E:Equilibria:Stability:solvability} in the range $0<\nu<10$ that is of interest, indicating that $\sigma$ does not change sign in this region (assuming $\sigma$ is continuous). Since $\sigma < 0$ somewhere in these regions (Figure~\ref{fig:Stability}b), we conclude that $\sigma < 0$ everywhere in these regions, i.e.~any equilibrium is stable.

For $\hyspare \gtrsim 0.1$, there are solutions of~\eqref{E:Equilibria:Stability:solvability} for $\nu < 10$ (Figure~\ref{fig:Stability}a), indicating that the growth rate $\sigma$ changes sign in these regions. For larger values of the channel bendability $\nu$, and volumes $V$,  (i.e. in the red regions of Figure~\ref{fig:Stability}a), equilibria have $\sigma < 0$, corresponding to stable equilibria (Figure~\ref{fig:Stability}b). For smaller values of $\nu$ and $V$ (in the blue regions, respectively), equilibria have $\sigma > 0$, and are unstable. It is perhaps surprising that the proportion of the equilibria that are unstable increases with the contact angle asymmetry $\hyspare$;  this can be rationalized by thinking again of $\hyspare$ as a geometric constraint: higher $\hyspare$ is associated with smaller $h(\xrighteq)$ (to maintain the ratio~\eqref{E:Equilibria:Eqs:hyspar_geometry}), and thus a larger change in the suction pressure when the droplet is perturbed (recall the suction pressure scales with the inverse of the channel width).

We stress that it is only with large values of $\hyspare$  that equilibria might be unstable. Since the results in this section are pertinent only for $\hyspare < \hysparmax$, these unstable equilibria are only possible for $\hysparmax \gtrsim 0.1$; with our typical receding angle $\theta_r = 0\si{\degree}$, this corresponds to a large contact angle difference of approximately $25\si{\degree}$. Moreover, the contact angle difference required to obtain a large hysteresis increases with larger $\theta_r$. In what follows, the results of this section are used to make predictions of the parameter values for which droplets are trapped; we shall consider only surfaces with $\hysparmax \lesssim 0.1$, for which any attainable equilibria are guaranteed by this analysis to be stable. 

\section{Droplet Trapping}\label{S:MobileDroplets}

Following the previous analysis describing when equilibria are possible, and assessing their stability, we are now in a position to describe the conditions under which droplets become trapped part way along the channel as a result of contact angle hysteresis. We have seen that, for $\hysparmax \lesssim 0.1$, droplets may become trapped in stable equilibria if they remain in the stage of the motion with $\xleft$ pinned; this, in turn, is possible, when  $\hysparmax>\hyspari$, i.e.~the contact angle asymmetry available is larger than that required to maintain the pinned state indefinitely. The crucial point to note is that if an equilibrium exists then the associated contact angle asymmetry $\hyspare(\nu, V, \xrighteq)\approx \hyspari(\nu, V, \xright^0 =\xrighteq)$: with given volume $V$ and bendability $\nu$, the contact angle asymmetry in equilibrium is approximately that for a pinned droplet with initial condition $\xright^0 = \xrighteq$. (Any difference between $\hyspare$ and $\hyspari$ is a result of the meniscus motion in the squeezing period, which is brief, making the difference relatively small, see Figure~\ref{fig:Numerics:ExampleTraces}.)  As an approximate criterion for the trapping of a droplet, therefore, we argue that droplets will be trapped if $\hysparmax \gtrsim \hyspare(\nu, V, \xrighteq = \xright^0)$, and will escape if $\hysparmax \lesssim\hyspare(\nu, V, \xrighteq = \xright^0)$.

\begin{figure}
\centering
\includegraphics[width=0.9\textwidth]{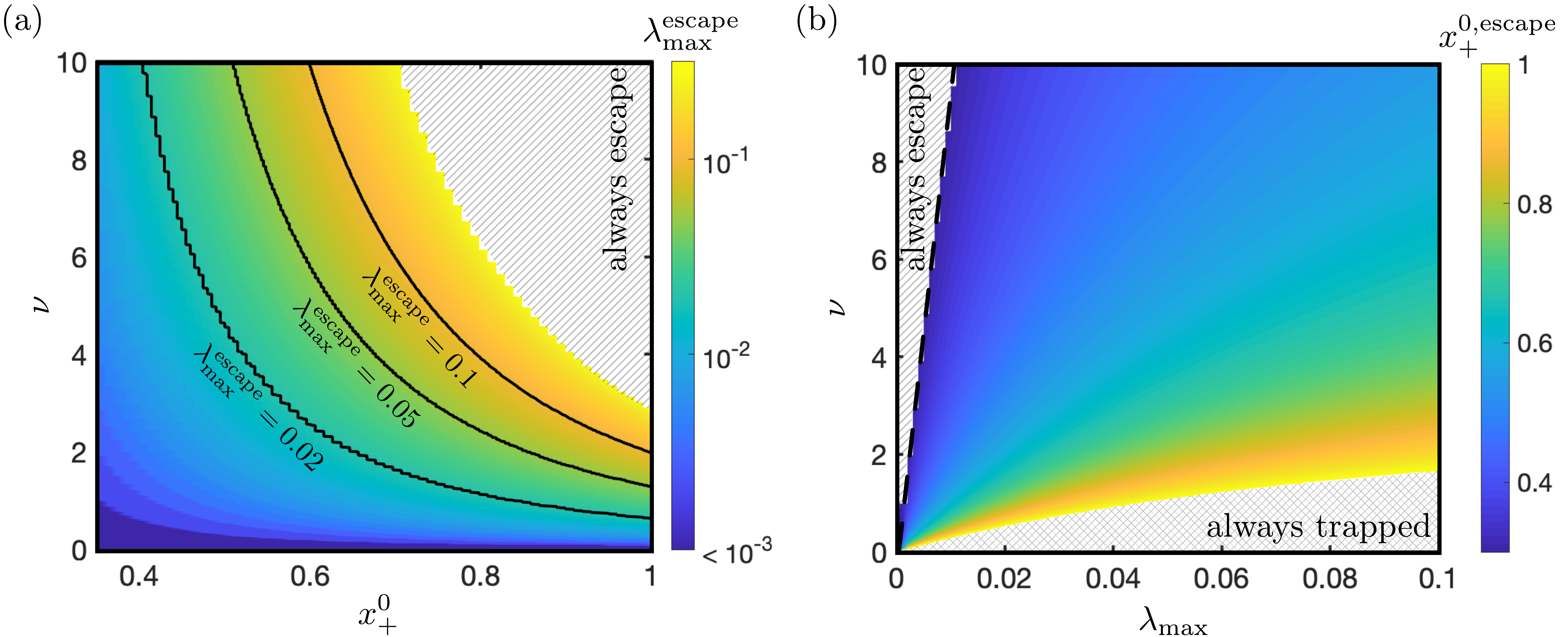}
\caption{Predictions from the equilibrium calculation of (a) $\hysparmax^{\text{escape}}$ (the largest value of $\hysparmax$ at which a droplet with initial position $\xright^0$ is able to escape) and (b) $\xright^{0, \text{escape}}$ (how far along the channel a droplet must start if it is to escape when the maximum contact angle asymmetry is $\hysparmax$). Both plots correspond to a droplet volume $V = 0.3$. In the upper right of (a) and upper left of (b), no equilibria with $ \hysparmax^{\text{escape}}= \hyspare $ exist (see Figure~\ref{fig:regime_diagrams_xright_and_hyspar_vs_nu}), and so we predict that droplets in configurations with $(\xright^0, \nu)$ that lie in this region will always escape, regardless of where they start in the channel. Similarly, configurations with $(\hysparmax, \nu)$ that lie in the hatched region in the lower right of (b) will always trap droplets of this volume. The black dashed curve in (b) indicates the prediction~\eqref{E:Trapping:always_escape_bdry} for the boundary of this `always escape' region. }\label{fig:escape_diagrams}
\end{figure}

With this approximate criterion, the equilibrium maps in Figure~\ref{fig:regime_diagrams_xright_and_hyspar_vs_nu} can be re-purposed as maps describing whether droplets will be trapped or not for a given value of $\hysparmax$ (these maps are shown again in Figure~\ref{fig:escape_diagrams} with updated labels to reflect this interpretation of the equilibria). Figure~\ref{fig:escape_diagrams}a shows the largest value of $\hysparmax$ at which a droplet of given $\xright^0$, $\nu$ and $V$ still ultimately escapes (as predicted by our approximate criterion); we denote this value by $\hysparmax^{\text{escape}}$. As we see from Figure~\ref{fig:escape_diagrams}a (and as was expected from the numerical solutions presented in \S\ref{S:Numerics}), $\hysparmax^{\text{escape}}$ is an increasing function of $\xright^0$: droplets that start closer to the free end are more likely to escape.  Moreover, for relatively low bendabilities, $\nu\lesssim 1$, droplets remain trapped wherever they start within the channel, even with very small hysteresis $\hysparmax\lesssim0.02$, which corresponds to an advancing contact angle $\theta_a \approx 11\si{\degree}$.

\begin{figure}
\centering
\includegraphics[width = 0.95\textwidth]{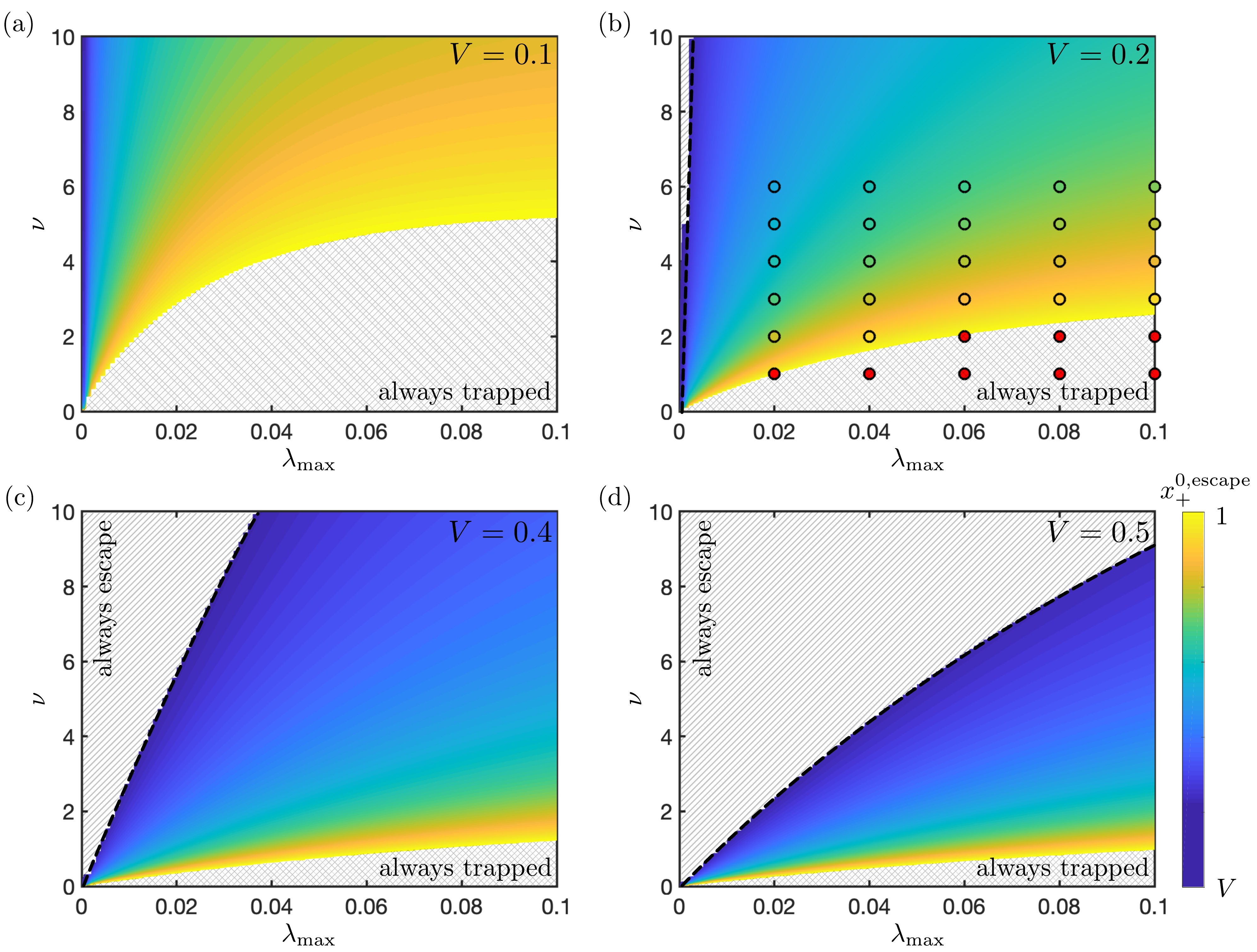}
\caption{Predictions from the equilibrium calculations of how far along the channel the droplet must start to escape, $\xright^{0,\text{escape}}$, if the maximum contact angle asymmetry is $\hysparmax$ and the bendability is $\nu$ (as in Figure~\ref{fig:escape_diagrams}) for droplet volumes $V=$ (a) 0.1, (b) 0.2, (c) 0.4, and (d) 0.5. The color-bar in (d) applies to each plot with the appropriate value of $V$.  The `always trapped' and `always escape' regions are as described in Figure~\ref{fig:escape_diagrams}. The filled circles in (b) correspond to the prediction of $\xright^{0,\text{escape}}$ based on numerical solutions of the full (dynamic) model (see main text).}\label{fig:escape_diagrams_variousV}
\end{figure}

Similarly, the regime diagram shown in Figure~\ref{fig:escape_diagrams}b  (in $(\hysparmax, \nu)$ space) can be interpreted as a map showing how far along the channel the initial position must be for the droplet to escape if the maximum contact angle asymmetry is $\hysparmax$; we denote this `escape position' by $\xright^{0,\text{escape}}$. (Another way to think of these data are as a surface separating trapped and escaping configurations: configurations with initial condition $\xright^0 < \xright^{0,\text{escape}}$ will be trapped, while those with initial condition $\xright^0 \geq \xright^{\text{escape}}$ will escape.) As expected, with larger $\hysparmax$ droplets need to start closer to the free end to escape. Similar maps for other droplet volumes $V = 0.1$, $0.2$, $0.4$, and $0.5$ are shown in Figure~\ref{fig:escape_diagrams_variousV}.

This equilibrium-based argument can only be used to predict $\xright^{0,\text{escape}}$ when such equilibria exist. If $(\nu, \hysparmax)$ are such that no equilibria exist (the hashed regions of Figure~\ref{fig:escape_diagrams}b), there are two possibilities:  there may be values of $\hyspar_e<\hysparmax$ for which there \emph{are} equilibria, in which case the droplets are always trapped, or there may be no equilibria for any $\hyspar_e<\hysparmax$, in which case they always escape.  The shape of these `always trapped' regions demonstrates that when surface tension is very weak (small $\nu$) only a small contact angle hysteresis $\hyspar_{\text{max}}$ is needed to ensure that droplets always get stuck, as we might expect. Moreover, the contact angle hysteresis needed to ensure that droplets are always trapped reduces for droplets of smaller volume, which are associated with smaller channel wall deflections (Figure~\ref{fig:escape_diagrams_variousV}).

The regions of parameter space in which droplets always escape only become appreciable for larger droplet volumes. (For example, for $V = 0.1$ this region is not clearly visible on the scale of Figure~\ref{fig:escape_diagrams_variousV}, but does exist.) The boundary between always escaping and sometimes being trapped corresponds to equilibria with $\xlefteq = 0$, whose location we expressed analytically in~\eqref{E:Equilibria:Eqs:analytic_sol_meniscus_position}--\eqref{E:Equilibria:Eqs:analytic_sol_open_ends}; we therefore predict that droplets will always escape when
\begin{equation}\label{E:Trapping:always_escape_bdry}
\nu > \frac{8}{V^4}\frac{\hyspar_{\text{max}}}{(1+\hyspar_{\text{max}})^2}\left(\frac{3\hyspar_{\text{max}} + 5}{5\hyspar_{\text{max}} + 5}\right)^4.
\end{equation}
The boundary between always escaping and some trapping, given by equality in~\eqref{E:Trapping:always_escape_bdry}, is included as the black dashed curves in Figure~\ref{fig:escape_diagrams}b and Figure~\ref{fig:escape_diagrams_variousV}. The sensitive dependence of~\eqref{E:Trapping:always_escape_bdry} on $V$, $\nu_{\mathrm{crit}}\propto V^{-4}$, elucidates why the always escape region is not resolved for smaller volume droplets. Note that for $\hysparmax \ll 1$, the criterion~\eqref{E:Trapping:always_escape_bdry} can be approximated by the simpler relation,
\begin{equation}\label{E:Trapping:always_escape_bdry_simple}
    \nu \gtrsim \frac{8}{V^4}\hysparmax.
\end{equation}
which agrees with~\eqref{E:Trapping:always_escape_bdry} to within 10\% for $\hysparmax = 0.05$ (corresponding to $\theta_a \approx 18\si{\degree}$), regardless of the value of $V$.

We conclude with a comparison between the results of our equilibrium-based predictions and numerical results of the full (dynamic) model. To do so, we compute $\xright^{0,\text{escape}}$ numerically using a bisection scheme, with the model equations solved numerically for many different initial conditions. We use $\xright^0 = 0.97$ as a first upper bound to avoid the situation where the `$+$' meniscus  is pushed onto the free end during the initial squeezing; droplets that are trapped even for $\xright^0 = 0.97$ are said to always be trapped. Similarly, we use $\xright^0 = V + 0.03$ (i.e.~ $\xleft^0 = 0.03$) as the first lower bound; droplets that escape even for $\xright^0 = V + 0.03$ are said to always escape. (The pattern of meniscus traces obtained in this way is qualitatively similar to those shown in Figure~\ref{fig:Numerics:initialpos}.) The values of $\xright^{0,\text{escape}}$ obtained numerically using this procedure agree well with the values determined from equilibrium calculation for $\hysparmax < 0.1$. This can be seen in Figure~\ref{fig:escape_diagrams_variousV}b, where exact agreement would be indicated by all coloured circles being indistinguishable from the background colouring used at that location; moreover, the red circles, which indicate parameter values for which droplets are never observed to escape lie exclusively within the empty region towards the right, where corresponding equilibria do not exist. We find that the numerically determined $\xright^{0,\text{escape}}$ are systematically lower than the equilibrium based predictions (although this is not clearly visible in Figure~\ref{fig:escape_diagrams_variousV}b), because the equilibrium calculation does not account for the meniscus motion in the squeezing period.

\section{Conclusions}\label{S:Conclusions}
In this paper, we have presented a theoretical analysis of the effect of contact angle hysteresis on the self-propulsion of droplets within deformable channels via \bendotaxis. We focused in particular on understanding when droplets may be unable to self-propel, and hence are trapped, by contact angle hysteresis. 

We developed a mathematical model in which contact angle hysteresis is parametrized  by the maximum contact angle asymmetry possible, $\hysparmax = \cos \theta_r / \cos \theta_a -1$. Numerical solutions of the model equations confirmed the intuition that when hysteresis is sufficiently strong ($\hysparmax$ sufficiently large), droplets may be trapped in equilibrium part way along the channel, but this scenario is only possible if droplets do not reach a translating stage defined by an advancing angle at the meniscus closest to the free end of the channel and a receding angle at the meniscus closest to the clamped end. By studying steady solutions of the model equations and assessing their linear stability, we determined that these equilibria are stable provided that the associated contact angle asymmetry is not too large, and focused on this case.

We identified the importance of the quantity $\hyspari$, the contact angle asymmetry required to hold a given droplet in the pinned state (when equilibria are possible); $\hyspari$ gives a simple criterion for whether a droplet will ultimately escape: droplets in channels with initial conditions such that the maximum contact angle asymmetry $\hysparmax \leq \hyspari$ will escape, while those with $ \hysparmax > \hyspari$ will not. In reality, $\hyspari$ can only be determined by a full dynamic simulation, but our analysis of equilibria gives an approximation for $\hyspari$, allowing us to re-purpose our regime diagrams of where equilibria exist to describe whether droplets of given parameters will be trapped or not. In doing so, we identified regions of parameter space in which droplets will always escape and other regions in which droplets are always trapped, regardless of where they start in the channel.  The shape of these regions are intuitive: when the channel bendability is small, only a small amount of contact angle hysteresis is required to trap droplets, and droplets are more likely to be trapped in channels with higher hysteresis (a prediction that is true even when the equilibrium analysis breaks down). 

\ab{Although we considered only wetting configurations here, we note that the main results are qualitatively similar for non-wetting configurations in which both the advancing and receding contact angles are greater than $90\si{\degree}$ (see Chapter 4 of ref.~\cite{Bradleyphdthesis}). The key quantitative differences are that, for a given droplet volume, the `always trapped' region is always larger, and the `always escape' region is always smaller, for non-wetting configurations than for wetting configurations. Briefly, the non-linearity in the Laplace pressure boundary condition is responsible: non-wetting droplets (which are associated with outwards channel wall deformations, $h > 1$) cannot create as large a droplet pressure, and thus deformation, while $\xleft$ is pinned, as wetting configurations (the meniscus pressure, which scaled with $1/h$ does not change as sharply when the meniscus advances into an outward tapered channel than when advancing into an inward tapered channel).}

Although our model is highly idealized, our results have implications for the exploitation of \ab{mechanisms that result in} self-propelling droplets. Most importantly, our results demonstrate that these \ab{such} systems are highly sensitive to contact angle hysteresis; in particular, we saw that even a modest amount of hysteresis is sufficient to trap droplets \ab{over} a wide range of parameter space and that the velocity of droplets are significantly reduced when contact angle hysteresis is present. For example, with $\nu V^4 = 10^{-2}$ and $\theta_r = 0$, corresponding to typical experimental values from~\cite{Bradley2019PRL}, we predict, using~\eqref{E:Trapping:always_escape_bdry_simple}, that droplets are only guaranteed to escape when $\hysparmax < 1.25\times 10^{-3}$, corresponding to a contact angle hysteresis of $\theta_a - \theta_r < 3\si{\degree}$.  This low value of the contact angle hysteresis that can be tolerated by \bendotaxis~suggests that low friction and hysteresis surfaces such as SLIPS~\citep{Wong2011Nature} (as used by~\citep{Bradley2019PRL}), or LIS~\citep{Solomon2014Langmuir} should be used to guarantee the success of \bendotaxis~as a mechanism for moving droplets.

The understanding of hysteresis that we have gained highlights the importance of minimizing it in applications in which droplet motion is desired. Such considerations may be particularly important for  natural examples of \bendotaxis. However, because these examples often occur on fibres (rather than in the channels considered here), the precise effect of hysteresis is likely to depend on the wettability of the drops involved. As an example in the non-wetting scenario, the spontaneous motion of condensed water drops out of the hairy texture on the legs of water striders helps to maintain a superhydrophobic state \cite{Wang2015PNAS}; our results lead us to expect that, in the presence of hysteresis, motion would  only occur once a sufficiently large droplet has condensed. As an example in the wetting scenario, small oil droplets on the barbules of bird feathers spread (causing the barbules to clump together) but larger drops move to the end and can be shaken off~\cite{Duprat2012Nature}. In this case, transitions between droplet shapes makes predicting the precise effect of hysteresis difficult,  though one might expect it to affect droplet motion  in each state, as well as the transitions between states. Going further, the understanding we have gained  may also open new opportunities for passive droplet control. In addition, there are several facets of the system that we have not considered, such as the clamping angle and variable bending stiffness of the channel walls, which may provide further opportunities for exploitation,  when combined with trapping by contact angle hysteresis.

\begin{acknowledgments}
This publication is based in part upon work supported by the European Research Council (ERC) under the European Union's Horizon 2020 research and innovation programme (grant agreement no.~637334, GADGET to D.V.), and the Leverhulme Trust (D.V.).
\end{acknowledgments}

\appendix
\section{Post-Translating Dynamics}\label{A:PostTranslatingDynamics}

\begin{figure}
    \centering
    \includegraphics[width=0.8\textwidth]{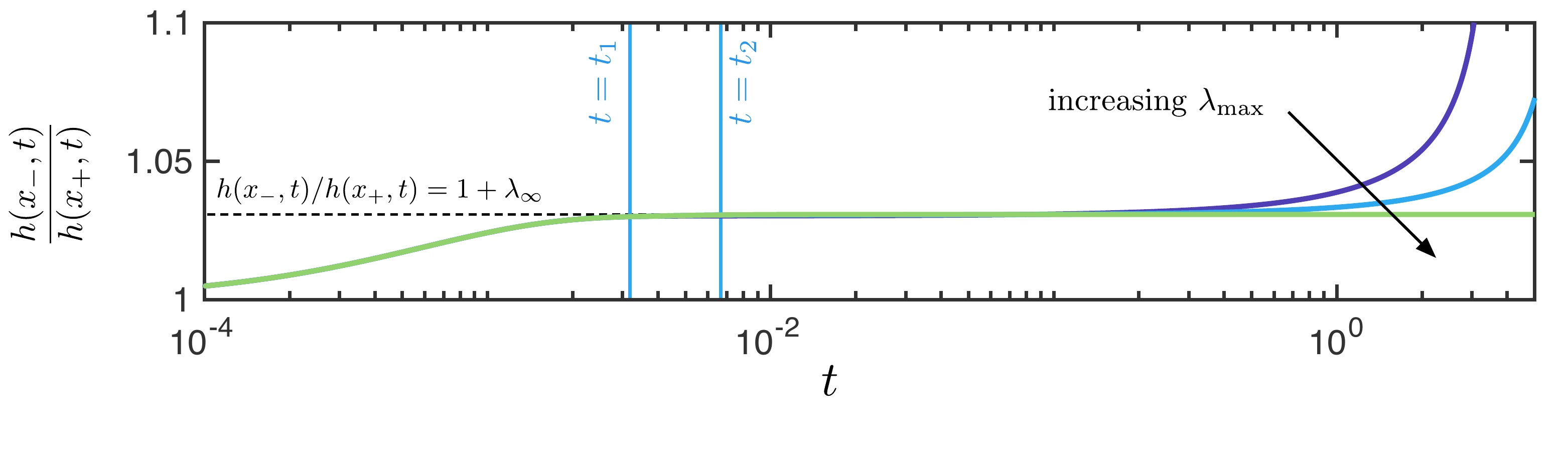}
    \caption{Evolution of the ratio of the channel widths at the menisci for the numerical solutions presented in Figure~\ref{fig:Numerics:ExampleTraces} (and described in \S\ref{S:Numerics:HysteresisDependence}), with $\nu = 4$, $\xright^0 = 0.65$, $V = 0.2$ and colour corresponding to the value of $\hysparmax$ as follows: $\hysparmax = 0$ (purple), $\hysparmax = 0.02$ (blue), and $\hysparmax = 0.04$ (green). The horizontal dashed lines indicate $h(\xleft,t)/h(\xright) = 1+ \hyspari$\ab{, the value in the trapped state, which is realized when $\hysparmax$ is sufficiently large.}}
    \label{fig:A:PostTranslatingDynamics}
\end{figure}

In this Appendix, we briefly mention the late time behaviour of \ab{droplet transport by} \bendotaxis, focusing in particular on the possibility that the `$-$' meniscus may be forced to pass through further pinning transitions, and justify \ab{our earlier assumption} that the droplet cannot be trapped during this period.

In \S\ref{S:Numerics}, we describe the behaviour of droplets up until they reach a translating stage (which they always reach provided that the contact angle hysteresis is sufficiently small) and move towards the free end of the channel. As the droplet approaches the free end of the channel, it may pass through another squeezing period (shown schematically in Figure~\ref{fig:Numerics:ExampleTraces}d): the high droplet pressure and distance from the clamped end results in large channel deformations that force the rear meniscus to decelerate again (by conservation of mass) until it becomes pinned and finally advances once more, back towards the clamped end of the channel (see Figure~\ref{fig:Numerics:ExampleTraces}d); all the while, the `$+$' meniscus continues to  accelerate and reaches the free end while the `$-$' meniscus is advancing. 

We justify ignoring the possibility of droplet trapping in this late period by referring to the condition~\eqref{E:Equilibria:Eqs:hyspar_geometry} that is necessary for equilibria.  \ab{Immediately as} the droplet reaches the translating stage,  \ab{the pressure gradient is zero at the `$-$' meniscus and negative at the (still advancing) `$+$' meniscus; the pressure is therefore more negative at the `$+$' meniscus, so $p(\xleft,t)/ p(\xright, t) < 1$, and therefore}
\begin{equation}\label{A:E:Inequality}
    \frac{h(\xleft, t)}{h(\xright,  t)} > \frac{h(\xleft, t)}{h(\xright,t)}\frac{p(\xleft,t)}{p(\xright,t)} = 1 + \hysparmax,
\end{equation}
\ab{where the equality comes from the Laplace pressure condition~\eqref{E:Model:NonDim:BCPressure}.}  \ab{As shown in Figure~\ref{fig:A:PostTranslatingDynamics}, the ratio $h(\xleft,t)/h(\xright,t)$ subsequently increases.} (This is because the inwards deformation of the channel walls lengthens the droplet and thus the relative stiffness of the channel walls, and thus the ratio of channel widths at the `$+$' and `$-$' menisci only increases.)  \ab{Hence, any equilibrium satisfying~\eqref{E:Equilibria:Eqs:hyspar_geometry} would have $\hyspare > \hysparmax$, which is not possible. }

\section{Locating Equilibria}\label{A:FindingEquilibria}

In this Appendix we describe the method used to find solutions of equations~\eqref{E:Equilibria:Eqs:Beam1}--\eqref{E:Equilibria:Eqs:Volume} describing an equilibrium configuration $h_e(x)$ whose menisci are located at $x = \xpmeq$.

We first `integrate out' the dry regions to give an equivalent problem defined only on the drop region $\xlefteq < x < \xrighteq$ with the effect of the dry regions encoded by effective boundary conditions. \ab{(}This procedure is described in detail in the Appendix of reference~\cite{Bradley2019PRL} \ab{ for the dynamic problem, but follows in the same way for the static problem considered here.) We find the } following system of equations:
\begin{equation}\label{A:E:FindingEquilibria:beam}
\dd{^4 h_e}{x^4} = p_0 \qquad \xlefteq < x< \xrighteq
\end{equation}
where 
\begin{equation}\label{A:E:FindingEquilibria:p0}
p_0 =- \frac{\nu}{h_e(\xrighteq)} = - \frac{\nu(\hyspar_e+1)}{h_e(\xlefteq)},
\end{equation} is the constant pressure within the droplet; the appropriate boundary conditions on the `wet' problem, i.e.~accounting for the behaviour in the dry regions,  are
\begin{align}
\left.\dd{^2 h_e}{x^2} - \frac{2}{\xlefteq^2}\left(2\xlefteq \dd{h_e}{x} - 3h_e+3\right)\right|_{x = \xlefteq} &= 0, \label{A:E:FindingEquilibria:EffBC_xl1}\\
\left.\dd{^3 h_e}{x^3} - \frac{6}{\xlefteq^3}\left(\xlefteq \dd{h_e}{x} - 2h_e+2\right)\right|_{x = \xlefteq} &= 0,\label{A:E:FindingEquilibria:EffBC_xl2}\\
\left.\dd{^2 h_e}{x^2} \right|_{x = \xrighteq}&=0\label{A:E:FindingEquilibria:EffBC_xu1}\\
\left.\dd{^3 h_e}{x^3} \right|_{x = \xrighteq}&=0,\label{A:E:FindingEquilibria:EffBC_xu2}
\end{align}
\ab{Equations~\eqref{A:E:FindingEquilibria:beam}--\eqref{A:E:FindingEquilibria:EffBC_xu2} must be solved} together with the volume constraint
\begin{equation}\label{A:E:FindingEquilibria:Volume}
V = \int_{\xlefteq}^{\xrighteq}h_e(x)~\mathrm{d}x.
\end{equation}

To \ab{make progress}, we first note that the channel shape in the drop region may be expressed as
\begin{equation}\label{A:E:FindingEquilibria:Shape}
h_e(x) = \frac{p_0}{24}(x - \xright)^4 + K_3 (x- \xright)^3 + K_2 (x- \xright)^2 + K_1 (x- \xright) + K_0
\end{equation}
where $K_i = K_i(\xrighteq, \xlefteq, p_0),~ i = 1,2,3,4$ \ab{are known}. Since the boundary conditions~\eqref{A:E:FindingEquilibria:EffBC_xu1} and~\eqref{A:E:FindingEquilibria:EffBC_xu2} are linear, the coefficients $K_i, ~i = 1,2,3,4$ are also linear in the equilibrium pressure $p_0$, and multinomials in $\xpmeq$. Using the solution~\eqref{A:E:FindingEquilibria:Shape},  the channel displacement at the menisci can then be expressed as
\begin{align}
h_e(\xrighteq) &= f_+(\xlefteq, \xrighteq) + p_0 g_+(\xlefteq, \xrighteq), \label{A:E:FindingEquilibria:meniscus_disp_1}\\
h_e(\xlefteq) &= f_-(\xlefteq, \xrighteq) + p_0 g_-(\xlefteq, \xrighteq)\label{A:E:FindingEquilibria:meniscus_disp_2}
\end{align}
where $f_{\pm}, g_{\pm}$ are known multinomials.

Inserting~\eqref{A:E:FindingEquilibria:meniscus_disp_1}--\eqref{A:E:FindingEquilibria:meniscus_disp_2} into the two pressure conditions~\eqref{A:E:FindingEquilibria:p0} gives two quadratic equations for the pressure $p_0$:
\begin{align}
p_0 \left[f_+(\xlefteq, \xrighteq) + p_0 g_+(\xlefteq, \xrighteq)\right]  &= -\nu, \label{A:E:FindingEquilibria:pressure_quadratic1}\\
 p_0 \left[f_-(\xlefteq, \xrighteq) + p_0 g_-(\xlefteq, \xrighteq)\right]  &= -\nu ( \hyspare+1).\label{A:E:FindingEquilibria:pressure_quadratic2}
\end{align}
Eliminating $p_0$ from~\eqref{A:E:FindingEquilibria:pressure_quadratic1}--\eqref{A:E:FindingEquilibria:pressure_quadratic2} gives a single multinomial whose coefficients depend on parameters $\nu, \hyspare$:
\begin{equation}\label{A:E:FindingEquilibria:master_quadratic}
F(\xlefteq, \xrighteq;\nu, \hyspare) = 0.
\end{equation}

\ab{For given $V, \nu$ and $\hyspar_e$, the equation~\eqref{A:E:FindingEquilibria:master_quadratic} is nonlinear and therefore expensive to solve numerically. It is more convenient to instead specify a meniscus position (typically $\xlefteq$) and solve for $\xrighteq$ only (thus yielding $h_e(x)$), and computing the associated volume $V$ a posteriori. }

\ab{Assuming that $\xlefteq$ is prescribed, \eqref{A:E:FindingEquilibria:master_quadratic} is simply a degree nine polynomial equation for $\xrighteq$. We solve this polynomial numerically using the \textsc{matlab} routine \texttt{roots}.  Once the equation~\eqref{A:E:FindingEquilibria:master_quadratic} has been solved for $\xrighteq$, we keep only those roots that correspond to physically relevant solutions --- i.e. those with $\xlefteq < \xrighteq < 1$ and for which the open end condition~\eqref{E:Equilibria:Eqs:OpenEnds} are satisfied. By sweeping over all permissible values of $\xlefteq$, we identify all possible equilibria.}

\bibliography{BHV21}% Produces the bibliography via BibTeX.

\end{document}